\documentclass{jfm}
\usepackage[usenames,dvipsnames]{xcolor}
\usepackage{commath}
\usepackage{natbib}
\usepackage{amsmath,amssymb,mathrsfs}
\usepackage{physics}
\usepackage{graphicx}
\usepackage{caption}
\usepackage{float}
\usepackage{subcaption}
\usepackage{tikz,pgfplots}
\usepackage[font=small,labelfont=bf]{caption}
\usepackage{newtxtext}
\usepackage{newtxmath}
\usepackage{hyperref}
\newcommand{\eq}[1]{equation \eqref{#1}}

\newcommand{\pard}[2]{\frac{\partial #1}{\partial #2}}

\newcommand{\beq}{\begin{equation}}
\newcommand{\eeq}{\end{equation}}
\newcommand{\oG}{G}
\newcommand{\goG}{{\mathscr G}}
\newcommand{\ooG}{{\mathbb G}}

\newcommand{\LL}{{\mathbb L}}
\newcommand{\A}{{\mathbb A}}
\newcommand{\eps}{\varepsilon}

\title{Transient Growth in Streaky Unbounded Shear Flow: A symbiosis of Orr and Push-over mechanisms}
\author{William Oxley\aff{1} \corresp{\email{woo21@cam.ac.uk}} and  Rich R. Kerswell\aff{1}}
\affiliation{\aff{1}DAMTP, Centre for Mathematical Sciences, Wilberforce Road, Cambridge CB3 0WA, UK}

\begin{document}
\maketitle

\begin{abstract}

Transient growth mechanisms operating on streaky shear flows are believed important for sustaining near-wall turbulence.
Of the three individual mechanisms present - Orr, lift-up and `push over' - Lozano-Duran et al. ({\em J. Fluid Mech.} {\bf 914}, A8, 2021) have recently observed that both Orr and push over need to be present to sustain turbulent fluctuations given streaky (streamwise-independent) base fields whereas lift-up does not. 
We show here, using Kelvin's model of unbounded constant shear augmented by spanwise-periodic streaks, that this is because the push-over mechanism can act in concert with a Orr mechanism based upon the streaks to produce much-enhanced transient growth. The model clarifies the transient growth mechanism originally found by Schoppa \& Hussain, ({\em J. Fluid Mech.} {\bf 453}, 57-108, 2002) and finds that this is one half of a linear instability mechanism centred at the spanwise inflexion points observed originally by Swearingen \& Blackwelder ({\em J. Fluid Mech.} {\bf 182}, 255-290, 1987). The instability and even transient growth acting on its own are found to have the correct nonlinear feedback to generate streamwise rolls which can then re-energise the assumed streaks through lift-up indicating a sustaining cycle.
Our results therefore support the view that while lift-up is believed central for the roll-to-streak regenerative process, it is Orr and push-over mechanisms that are {\em both} key for the streak-to-roll regenerative process in near-wall turbulence. 

\end{abstract}

%
%
\section{Introduction} \label{intro}

Efforts to understand wall-bounded turbulence have naturally focussed on the wall and the (coherent) structures which form there \citep{Richardson22}.
The consensus is that there is (at least) a near-wall sustaining cycle \citep{Hamilton95,  Waleffe97, JimenezPinelli99, FarrellIoannou12} involving predominantly streaks and streamwise rolls (or vortices) which helps maintain the turbulence \citep[e.g. see the reviews][]{Robinson91, Panton01, Smits11, Jimenez12, Jimenez18}. The generation of these streaks from the rolls is commonly explained  by the (linear) transient growth `lift-up' mechanism \citep{Ell-75,Landahl80}, but how rolls are regenerated from the streaks has proven a little less clear due to the need to invoke nonlinearity at some point.

Just focusing on the initial linear part, \cite{Sch-02} suggested that transient growth mechanisms on the streaks were actually more important than (linear) streak instabilities, and that it was these transiently growing perturbations which fed back to create streaks through their nonlinear interaction. While this view has been contested \citep[e.g.][]{Hoepffner05,Cassinelli17, Jimenez18}, it is supported by recent cause-and-effect numerical experiments by \cite{Loz-21} who looked more closely at all the linear processes present. In particular, \cite{Loz-21} isolated the influence of the three different transient growth mechanisms: the familiar Orr \citep{Orr-07} and lift-up \citep{Ell-75} mechanisms present for a 1D shear profile $U(y)$ and a far less-studied `push-over' mechanism which can only operate when the base profile has spanwise shear i.e. $U(y,z)$.  They found, somewhat unexpectedly, that the Orr and push-over mechanisms are {\em both} essential to maintain near-wall turbulence when streaky base flows were prescribed but lift-up is not: see their \S6.4 and figure 24(a). 

\cite{Mar-24} investigated this further by looking at  the transient growth possible on a wall-normal shear plus monochromatic streak field consistent with the buffer region at the wall. Over appropriately short times (e.g. one eddy turnover time as proposed by \cite{But-93}), they found a similarly clear signal that lift-up is unimportant whereas the removal of push-over dramatically reduced the growth: see their figure 7. The necessity to have push-over operating with the Orr mechanism indicates they are working symbiotically. How this happens, however, is puzzling from the timescale perspective as Orr is considered a `fast' mechanism which operates over inertial timescales whereas push-over looks a `slow' mechanism operating over viscous timescales. This latter characterisation comes from an analogy with lift-up in which viscously-decaying wall-normal velocities (as present in streamwise rolls) advect the base shear to produce streaks. Push-over (a term coined by \cite{Loz-21}) similarly involves viscously-decaying spanwise velocities advecting the spanwise streak shear. Understanding exactly how these two mechanisms constructively interact is therefore an interesting issue.

%
%
\begin{figure}
\centering
\includegraphics[width=0.5\linewidth]{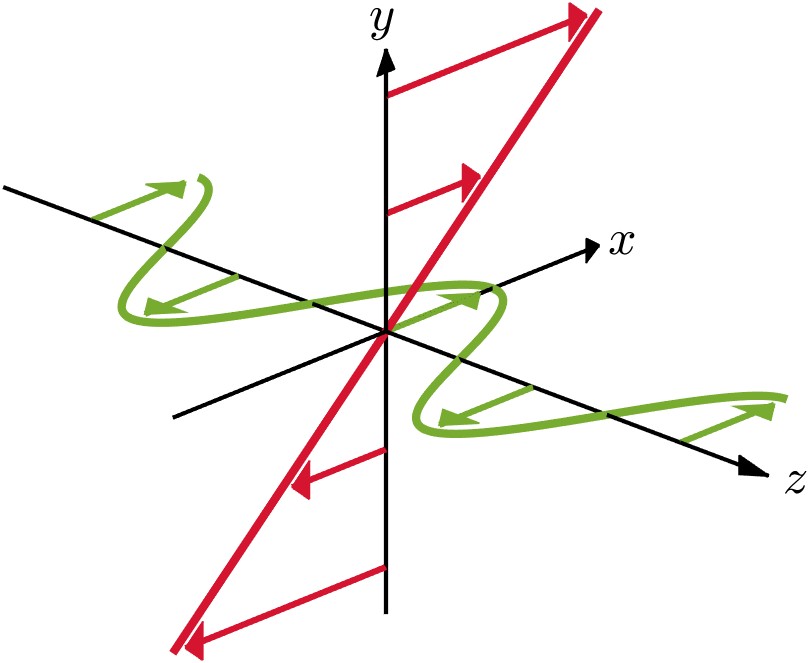}  
\caption{The streaky unbounded shear flow studied here: see \eq{UB}. The (green) spanwise streaks extend Kelvin's (red) popular constant shear model. }
\label{base}
\end{figure}

The purpose of this paper is to lay bare this interaction by exploring it in an augmented version of Kelvin's famous constant shear model \citep{Kel-87}.  This simple model - an unbounded shear ${\bf U}=y {\bf \hat{x}}$ (just the red flow in figure \ref{base}) - was used by \cite{Orr-07} for his seminal work and has been important in clarifying the characteristics  of both Orr and lift-up mechanisms subsequently \citep[e.g.][]{Far-93,Jim-13,Jiao-21} and as a shear-flow testbed otherwise \citep[e.g.][]{Moffatt67, Marcus77}. The key features of the model are that the base flow is: 1. unbounded and so  not restricted by any boundary conditions; and 2. a linear function of space. 
These together permit plane wave  solutions  to the perturbation evolution equations where the spatially-varying base advection can be accounted for by time-dependent wavenumbers. This leaves just 2 ordinary differential equations (ODEs) for the cross-shear velocity and cross-shear vorticity to be integrated forward in time. These `Kelvin' modes form a complete set but, unusually, are not individually separable in space and time 
and so the representation differs from the usual plane wave approach with constant wavenumbers. Kelvin and Orr used these modes to study unidirectional shear but they can also be used more generally to study the stability of time-dependent spatially-linear flows \citep{Craik86,Craik89}, flows with closed streamlines such as elliptical \citep{Bayly86, Landman87, Waleffe90} and precessing flows \citep{Kerswell93}, or flows with more physics included such as stratification \citep{Hartman75, Miyazaki92}, rotation \citep{Tung83, Leblanc97, Salhi10}, magnetic fields \citep{Craik88} or elasticity \citep{Lagnado85}.

The augmented base flow considered here - shown in Figure \ref{base} and equation (\ref{UB}) below - builds in a streak field  which introduces  spatially-periodic spanwise shear. This is now not purely linear in space and so a Kelvin mode is no longer a solution of the linearised perturbation equations. Instead, a single sum of Kelvin modes over spanwise wavenumbers is needed, but, importantly, the wall-normal shear can be handled as usual, removing the unbounded advective term from the system. This means the model system is still a very accessible `sandbox' in which to study the transient growth mechanisms of Orr, lift-up and now, crucially, also `push-over'. The price to be paid for introducing the streak field is an order of magnitude increase in the number of ODEs to be solved, but, since this is increased from 2 to $O(20)$, it is trivial by today's standards.

The plan of the paper is as follows. Section \ref{ProbForm} introduces the model, the evolution equations and discusses appropriate parameter values. Section \ref{nostreak} revisits Kelvin's unbounded constant-shear model, presenting some new  large-$\Rey$ asymptotic scaling laws and discussing the timescales for Orr and lift-up growth mechanisms.  The presence of streaks is introduced in  \S4, with the 2D limit of no streamwise variation used in \S \ref{2DTG} to illustrate how the push-over mechanism behaves when it acts alone. This is followed by a general analysis of the transient growth possible for the full 3D system in \S \ref{3DTG} which is found to clearly capture  the symbiotic relationship between Orr and push-over. Section \ref{simpmodsec} then describes a severely truncated 2-variable system which contains the essence of  how Orr and push-over help each other to generate enhanced growth.  \S \ref{4.4} and \ref{4.5} discuss the energy growth possible in this system which is centred on a linear instability mechanism. \S \ref{4.6} discusses how this linear instability becomes only transient when diffusion is reinstated and derives some simple estimates for the transient energy growth possible which match well with what is found in the full 3D system. \S \ref{4.7} shows that the 2-variable system does not nonlinearly drive any streamwise rolls and so motivates examination of a  more realistic 3-variable system in \S \ref{4.8} which does. \S \ref{4.9} explains why lift up hampers the enhanced growth produced by push over and Orr working together and \S \ref{4.10} relates the results to previous work considering the full Navier-Stokes equations in a realistic geometry.
A final discussion follows in \S \ref{disc}.

%
%
\section{Formulation} \label{ProbForm}

%
%
\subsection{Governing Equations}

We consider a unidirectional base flow velocity in the $x$-direction which has a constant shear in $y$ and a spatially periodic `streak' shear varying in the spanwise $z$-direction,
\begin{equation}
\boldsymbol U_B = U_B(y,z)\boldsymbol e_x = [y + \beta \cos{(k_zz)}] \boldsymbol e_x,
\label{UB}
\end{equation}
where $\beta$ is the dimensionless streak strength: see Figure \ref{base}. The system has been non-dimensionalised by the $y$-shear rate, $S$, and the initial $k_y$ wavenumber (i.e. by a lengthscale $L:=L_y/2\upi$ where $L_y$ is the initial perturbation wavelength in $y$), so that the streak wavenumber $k_z$ is a parameter of the problem. This allows straightforward access to the well-studied Kelvin problem of $\beta \rightarrow 0$ while using  the spanwise wavenumber as an inverse length scale does not.

The Navier-Stokes equations linearised around $\boldsymbol U_B$ for a perturbation ${\boldsymbol u}:=(u,v,w)$ and associated pressure perturbation $p$ are then
\begin{gather}
\pard{\boldsymbol u}{t} + [y + \beta \cos{(k_zz)}] \pard{\boldsymbol u}{x} + [v-\beta w k_z \sin{(k_zz)}]  \boldsymbol e_x + \nabla p = \frac{1}{\Rey} \Delta \boldsymbol u,  \label{G1}\\
\boldsymbol \nabla \boldsymbol \cdot \boldsymbol u = 0. \label{G2}
\end{gather}
Here, the Reynolds number is $\Rey:=L^2 S/\nu$, where $\nu$ is the kinematic viscosity.
Taking    
$\boldsymbol e_y \! \cdot \! \boldsymbol \nabla \times \boldsymbol \nabla \times$ 
and
$\boldsymbol e_y \!\cdot\! \boldsymbol \nabla \times$ 
of (\ref{G1}) leads to a pair of equations for $v$ and the $y$-shearwise vorticity  $\eta:= \partial u / \partial z - \partial w/\partial x$ respectively,
\begin{align}
\bigg[\pard{}{t} + [y + \beta \cos{(k_zz)}]  \pard{}{x}  - \frac{1}{\Rey}\Delta\bigg] & \Delta v 
+  2\beta k_z \sin{(k_z z)} \left[\frac{\partial^2 w}{\partial x \partial y} - \frac{\partial^2 v}{\partial x \partial z} \right]
\nonumber\\
&\hspace{3.5cm}- \beta k_z^2 \cos{(k_zz)}\pard{v}{x} =0, \label{Gveta1} \\
\bigg[\pard{}{t} + [y + \beta \cos{(k_zz)}] \pard{}{x}  - \frac{1}{\Rey}\Delta\bigg] & \eta 
+ \pard{v}{z} + \beta k_z \sin{(k_zz)}\pard{v}{y} - \beta w k_z^2 \cos{(k_zz)} =0.
\label{Gveta2}
\end{align}

%
%
\subsection{Kelvin Modes} \label{KM}

The well-known trick to handle the $y$-dependent advection term is to allow the wavenumbers of the perturbation to be time-dependent \citep{Kel-87}
giving a `Kelvin' mode
\begin{equation}
[u,v,w,p,\eta](x,y,z,t) = [\hat{u},\hat{v},\hat{w},\hat{p},\hat{\eta}](t)\mathrm{e}^{\mathrm{i}\left[k_x x+(1-k_x t)y+k_z z\right]}
\label{KW}
\end{equation}
which is actually a full solution to the unlinearised perturbation equations since ${\bf k} \cdot \boldsymbol u=0$ by incompressibility \citep[e.g.][]{Craik86}. For $\beta \neq 0$, the extra streak shear unavoidably couples different spanwise wavenumbers and the perturbation ansatz has to be extended to 
\begin{equation}
[u,v,w,p,\eta](x,y,z,t) = \sum_{m=-M}^M [\hat{u}_m,\hat{v}_m,\hat{w}_m,\hat{p}_m,\hat{\eta}_m](t)\mathrm{e}^{\mathrm{i}\left[k_xx+(1-k_xt)y+mk_z z\right]},
\label{KWs}
\end{equation}
where, formally, $M$ should be infinite but, practically, is chosen large enough so as not to influence the results. This ansatz could be extended further by shifting the spanwise wavenumbers by a fraction of the base wavenumber - so $mk_z \rightarrow mk_z+\mu$ where $-\tfrac{1}{2}k_z \leq \mu< \tfrac{1}{2}k_z$ is the modulation parameter in Floquet theory - but only $\mu=0$ is treated here so the perturbation has the same wavelength as the base field (i.e. there is no modulation) consistent with numerical observations e.g. \cite{Sch-02}. Also, since $\boldsymbol U_B$ is symmetric about $z=0$, perturbations which are either symmetric (`varicose')  or antisymmetric (`sinuous')  in $u$ about $z=0$ can be pursued separately. However, the computations are sufficiently low-cost that it was easier to just consider both cases together.

The continuity equation and definition of $\eta$ make it possible to express $\hat{u}_m$ and $\hat{w}_m$ in terms of  $\hat{v}_m$ and $\hat{\eta}_m$ as follows
\begin{equation}
\hat{u}_m = \frac{-k_x(1-k_xt)\hat{v}_m - \mathrm{i}mk_z\hat{\eta}_m}{k_x^2+m^2 k_z^2} \quad \text{and} \quad \hat{w}_m = \frac{-m k_z(1-k_xt)\hat{v}_m + \mathrm{i}k_x\hat{\eta}_m}{k_x^2+ m^2 k_z^2}, \label{uweq}
\end{equation}
and then equations (\ref{Gveta1}) and (\ref{Gveta2}) require for each integer $m$:
\begin{align}
\dot{\hat{v}}_m +\left[-\frac{2k_x(1-k_xt)}{k_m^2}+ \frac{k_m^2}{\Rey}\right]\hat{v}_m &= \frac{\mathrm{i}\beta k_x}{2k_m^2}\left[ 2(m+1)\frac{k_z^2k_{m+1}^2}{h_{m+1}^2} - k_z^2 - k_{m+1}^2 \right]\hat{v}_{m+1} & \nonumber\\ 
&-\frac{\mathrm{i}\beta k_x}{2k_m^2}\left[ 2(m-1)\frac{k_z^2k_{m-1}^2}{h_{m-1}^2} + k_z^2 + k_{m-1}^2 \right]\hat{v}_{m-1}& \nonumber\\ 
&+ \frac{\beta k_x^2 k_z (1-k_xt)}{k_m^2h_{m+1}^2}\hat{\eta}_{m+1}  - \frac{\beta k_x^2 k_z (1-k_xt)}{k_m^2h_{m-1}^2}\hat{\eta}_{m-1}& \label{dv}
\end{align}
and
\begin{align}
\hspace{1.5cm} \dot{\hat{\eta}}_m +\frac{k_m^2}{\Rey}\hat{\eta}_m = 
 - &\frac{\mathrm{i}k_x\beta}{2}\left(1- \frac{k_z^2}{h_{m+1}^2} \right)\hat{\eta}_{m+1} 
 - \frac{\mathrm{i}k_x\beta}{2}\left(1-\frac{k_z^2}{h_{m-1}^2} \right)\hat{\eta}_{m-1} \nonumber \\
&-  \mathrm{i} m k_z\hat{v}_m  
- \frac{\beta k_z(1-k_xt)}{2}\left(\frac{(m+1)k_z^2}{h_{m+1}^2}-1\right)\hat{v}_{m+1}  \nonumber \\
&\hspace{1.6cm} - \frac{\beta k_z (1-k_xt)}{2}\left(\frac{(m-1)k_z^2}{h_{m-1}^2}+1\right)\hat{v}_{m-1}, \label{deta}
\end{align}
where $h_m^2:=k_x^2+m^2 k_z^2$ and $k_m^2:= k_x^2+(1-k_xt)^2+m^2 k_z^2$.  These $2(2M+1)$  coupled complex ODEs are readily integrated forward in time from given initial conditions using MATLAB's ode45. 

%
%
\subsection{Kinetic Energy and Optimal Gain}

The volume-averaged kinetic energy is defined as
\begin{equation}
E(t) := \frac{1}{V_\Omega}\int_\Omega \frac{1}{2} |{\boldsymbol u({\bf x},t})|^2 d\Omega 
=\frac{1}{2} \sum_{m=-M}^M \frac{1}{h_m^2} \left[k^2_m\abs{\hat{v}_m}^2 +\abs{\hat{\eta}_m}^2\right],
\label{energy}
\end{equation}
where $\Omega:=[0,2\upi/k_x] \times [0,2 \upi] \times [0,2\upi/k_z]$ and $V_\Omega$ is the volume of $\Omega$. In this paper we define the `optimal gain' as the largest value of $E(T)/E(0)$ over all possible initial conditions, 
%
%
\begin{equation}
\oG(T,k_x,k_z;\beta,\Rey) := \max_{{\bf u}(0)}\frac{E(T,k_x,k_z;\beta,\Rey)}{E(0,k_x,k_z;\beta,\Rey)}.
\label{optgain}
\end{equation}
%
%
The `global optimal gain' will be the result of optimising this quantity over the wavenumbers,
\begin{equation}
\goG(T;\beta,\Rey) := \max_{\{k_x,k_z\}}\oG(T,k_x,k_z;\beta,\Rey),
\label{optgainglobal}
\end{equation}
and the `overall optimal gain' will be the result of optimising this over the time $T$,
%
%
\begin{equation}
\ooG(\beta,\Rey) := \max_{T} \goG (T;\beta,\Rey).
\label{optgainglobal}
\end{equation}

%
%
\subsection{Parameter Choices: buffer layer estimates} \label{paramsec}

The focus in this study is to consider values for the five physical parameters - $T,k_x,k_x,\beta$ and $\Rey$ - appropriate for the buffer layer \citep[e.g.][]{Jimenez18,Loz-21}. Given there is no simple characterisation of the mean flow profile here as it adjusts from the viscous sublayer to the log layer and only order of magnitude estimates are sought, we adopt the empirical log layer representation
\beq
U^+(y^+) = U/u_\tau\approx \frac{1}{\kappa} \ln y^+ +B,
\eeq
(where $y^+=yu_\tau/\nu$ is the distance from the wall in so-called viscous units, $u_\tau$ is the friction velocity at the wall, $\kappa \approx 0.4$ and $B \approx 5$), even in the buffer region. Then at a typical buffer scale of $y^+=20$, this gives the local Reynolds number 
\beq
\Rey_{\text{local}} \approx y^+ \biggl( \frac{1}{\kappa}\ln y^+ +B \biggr) \approx 250
\eeq
which is large but not very large.

%
%
A representative $T$ can be found using the time quoted by \cite{Loz-21} for the maximum gain seen in numerical computations, which is $t=0.35h/u_\tau$ ($h$ being the half-channel width) at $\Rey_{\tau}:=hu_\tau/\nu=180$ and comparable with the eddy turnover time at the appropriate distance from the wall \citep{Mar-24}. The local shear rate at $y^+$ using the log layer approximation is $S=dU/dy=\Rey_\tau u_\tau/h\, dU^+/dy^+=\Rey_\tau u_\tau/(h \kappa y^+)$ which gives a time in our inverse shear rate units of
\beq
T \sim 0.35 \frac{h}{u_\tau} S= 0.35 \frac{\Rey_\tau}{\kappa y^+} \approx 8
\eeq
for $y^+=20$. 

%
%
In terms of wavenumbers, we take the initial  $k_y$ (cross-shear) wavenumber used to nondimensionalise the model as $2\upi/y^+$ (i.e. the wavelength is the distance to the wall), so the other non-dimensionalised wavenumbers are  $k_z \sim O( y^+/(\lambda^+_z \approx 100) )$ and $k_x \sim O( y^+/(\lambda^+_x \approx 300) )$. Therefore both wavenumbers are $O(0.1$-$1)$.

%
%
The magnitude of $\beta$ can be estimated by comparing shear rates. Typical root-mean-square velocities near the wall are $u_{rms} \approx 1$ (in units of $u_\tau$) which can be used to estimate typical streak amplitudes. Comparing the ratio of streak shear to wall-normal shear with the model (\ref{UB}) gives 
\beq
\frac{k_z u_{rms}}{1/(\kappa y^+)} \approx \frac{k_z\beta/\sqrt{2}}{1} 
\eeq
or $\beta \approx 11$ at $y^+=20$ (the $\sqrt{2}$ factor compensates for the cosine dependence in (\ref{UB})).  Given this, we consider values of $\beta$ from 1 to 5 in the below (going to higher $\beta$ causes problems with numerical overflow in our model: see figure \ref{VaryBeta}).

%
%

\section{Unbounded Constant Shear Flow ($\beta=0$): Kelvin's problem} \label{nostreak}

Removing the streaks ($\beta=0$) recovers Kelvin's classical unbounded constant shear model \citep{Kel-87} which ever since has proved a popular testing ground for ideas \citep[e.g.][]{Orr-07,Moffatt67,Craik86,Far-93,Jim-13,Jiao-21}.  The evolution equations are simply
\begin{equation}
(k^2\hat{v})_t = - \frac{k^2}{\Rey}(k^2\hat{v}), \qquad \hat{\eta}_t = - \frac{k^2}{\Rey}\hat{\eta}  -  \mathrm{i}k_z\hat{v},
\label{NoStreak1}
\end{equation}
with the solution 
\begin{equation}
\hat{v}(t) = \frac{k^2(0)}{k^2(t)}\hat{v}(0)\mathrm{e}^{\phi(t)}, \qquad \hat{\eta}(t) = \left\{\frac{-\mathrm{i} k_z k^2(0) \hat{v}(0)}{k_x h}\left[\theta(t)-\theta(0) \right] + \hat{\eta}(0)\right\}\mathrm{e}^{\phi(t)},
\label{NoStreak2}
\end{equation}
where 
\begin{equation}
\phi(t) := -\frac{1}{\Rey} \int_0^t k^2(\tau) \, d\tau, \quad \text{and} \quad 
\theta(t) := \tan^{-1}\left[\frac{h}{1-k_x t}\right] \label{NoStreak3}
\end{equation}
\citep{Far-93,Jiao-21}. It is well known that two growth mechanisms are present which are revealed by taking the appropriate 2-dimensional special case: $k_z=0$ possesses the Orr mechanism \citep{Orr-07} only and $k_x=0$ possesses the lift-up mechanism \citep{Ell-75} only.

%
%
\subsection{Orr mechanism}

With $k_z=0$, the shearwise vorticity can only exponentially decay leaving the shrinking of $k^2(t):=k_x^2+(1-k_xt)^2$ with time in the expression for $\hat{v}(t)$ - see (\ref{NoStreak2}) - the sole growth mechanism. Growth occurs for $t\leq 1/k_x$ with the viscous (exponential) damping term moderating this. The fact that the growth time does not scale with viscosity means the Orr mechanism is inertial and so considered `fast'. 

For large $\Rey$ and $1\lesssim \! T \!= \! o(\Rey)$, the viscous damping can be ignored and then $k_x=\tfrac{1}{2}( \sqrt{T^2+4}-T)$ maximises the global optimal growth at  
\begin{equation}
\goG^{\text{Orr}} =1/k_x^2 \sim T^2
\label{Orr_T_1}
\end{equation}
for $T^2 \gg 4$. For $1 \ll T=O(\Rey)$, the optimal wavenumber is still related to the time via $k_x=1/T$ to leading order and it is simple to show that the overall growth maximum over $T$ (the overall global optimal) is 
\beq
\ooG^{\text{Orr}}= \frac{9}{\mathrm{e}^2}\Rey^2, \quad T^{\text{Orr}}= 3\Rey, \quad k_x^{\text{Orr}}= \frac{1}{3 \Rey}, \label{Orr_asym}
\eeq
where $\mathrm{e}=\exp(1)=2.7183$. So, somewhat paradoxically, the maximum growth over all times produced by the `fast' Orr mechanism actually occurs on a `slow' viscous time scale.

%
%
%
\subsection{Lift-Up mechanism}

With $k_x=0$, there is no growth possible in $\hat{v}$ but this velocity component forces  the shearwise vorticity $\hat{\eta}$ which grows. This growth ultimately is turned off by the viscous decay of $\hat{v}$ and so lift-up is viewed as a `slow' viscous process. The overall optimal lift-up growth ($k_x=0$) is, 
\beq
\ooG^{\text{lift-up}}= \frac{4}{27e^2}\Rey^2, \quad T^{\text{lift-up}}= \frac{2}{3}\Rey, \quad k_z^{\text{lift-up}}= \frac{1}{\sqrt{2}}, \label{LiftUp_asym}
\eeq
which is consistent with this. Here, since $1\ll T=O(\Rey)$, the initial shearwise vorticity can be ignored facilitating the derivation of (\ref{LiftUp_asym}) while for $1 \lesssim T=o(\Rey)$, this is no longer valid.

%
%
\begin{figure}
\centering
\includegraphics[width=0.75\linewidth]{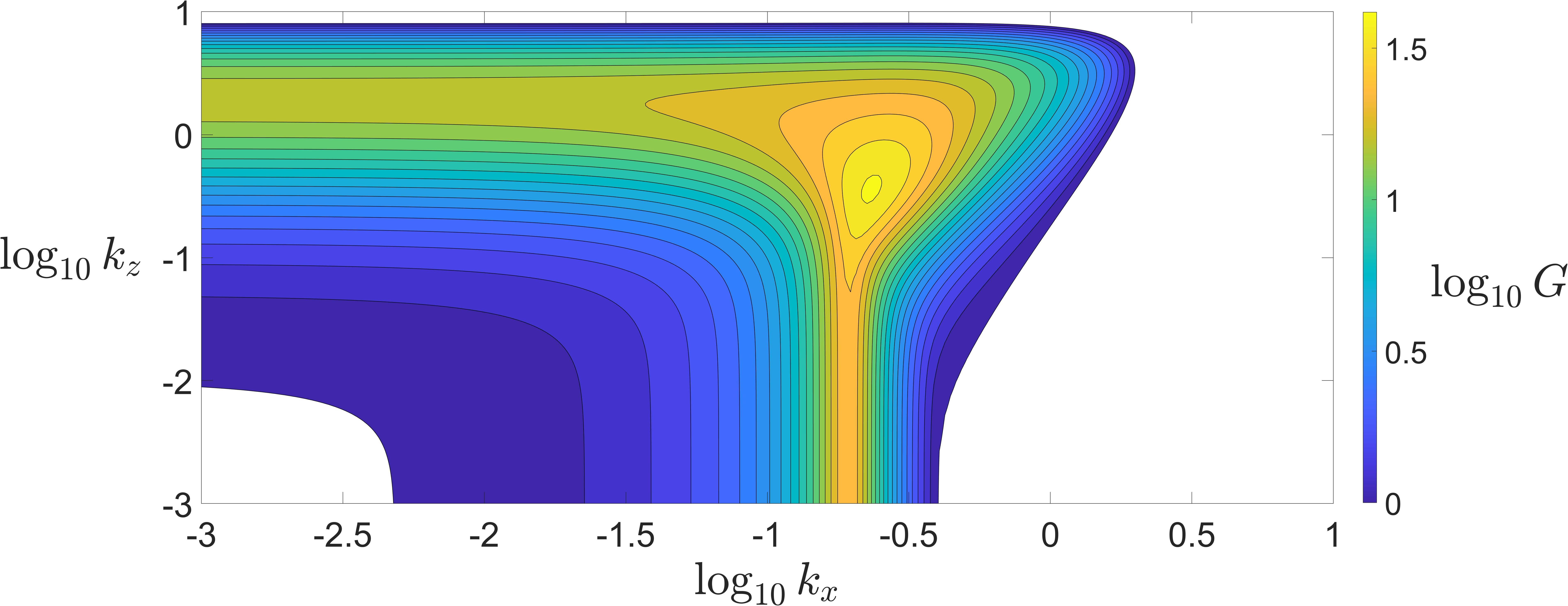}  
\caption{A contour plot of optimal growth $G$ in wavenumber space for $T=5$ and $\Rey=200$. This figure shows that the optimal early time gain occurs for a three-dimensional perturbation, at a location in wavenumber space close to the $k_x$ value associated with the maximum gain possible through the Orr mechanism. }
\label{EarlyNoStreakFig}
\end{figure}

%
\begin{figure}
\centering
\includegraphics[width=0.40\linewidth]{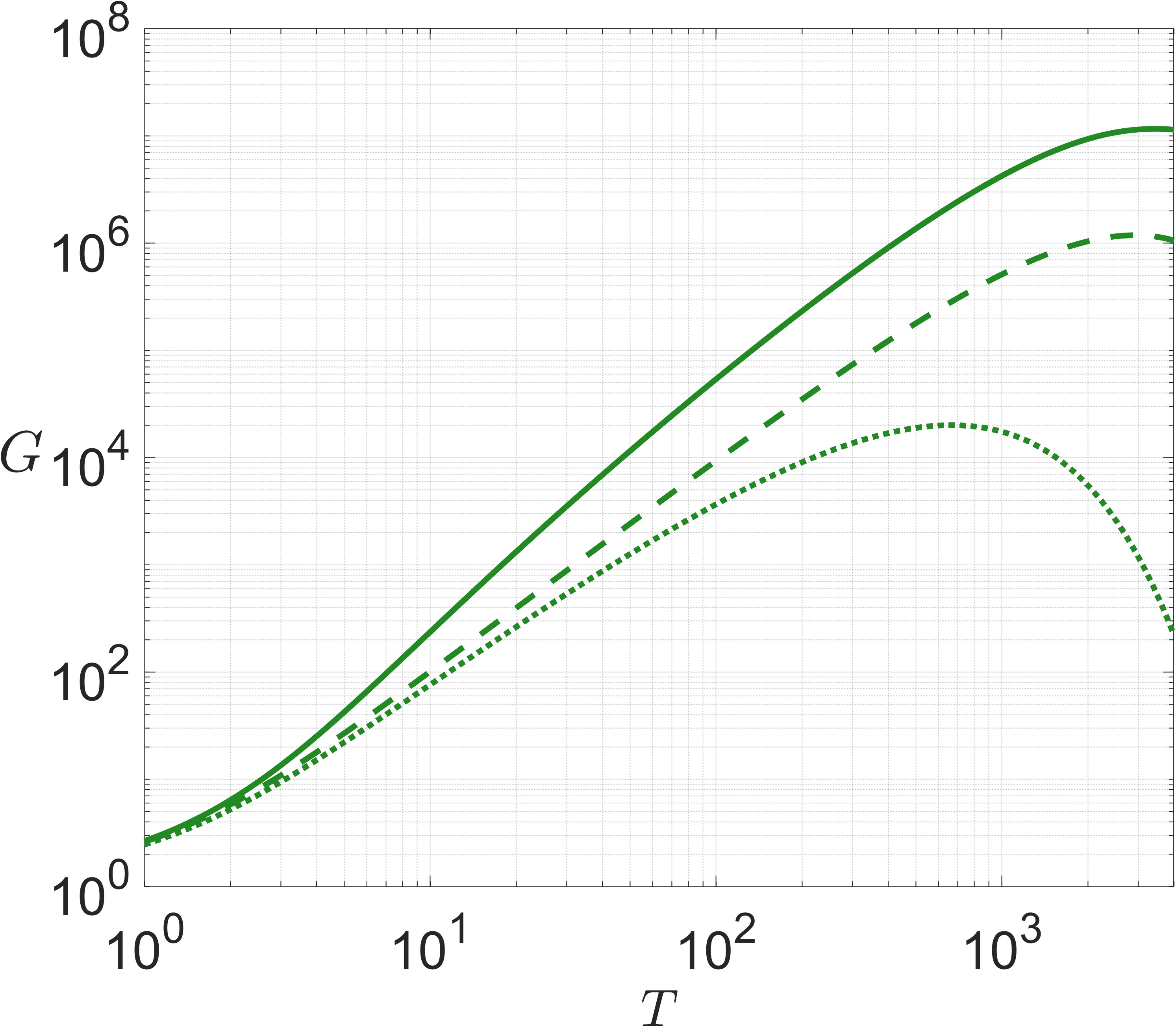}
\includegraphics[width=0.46\linewidth]{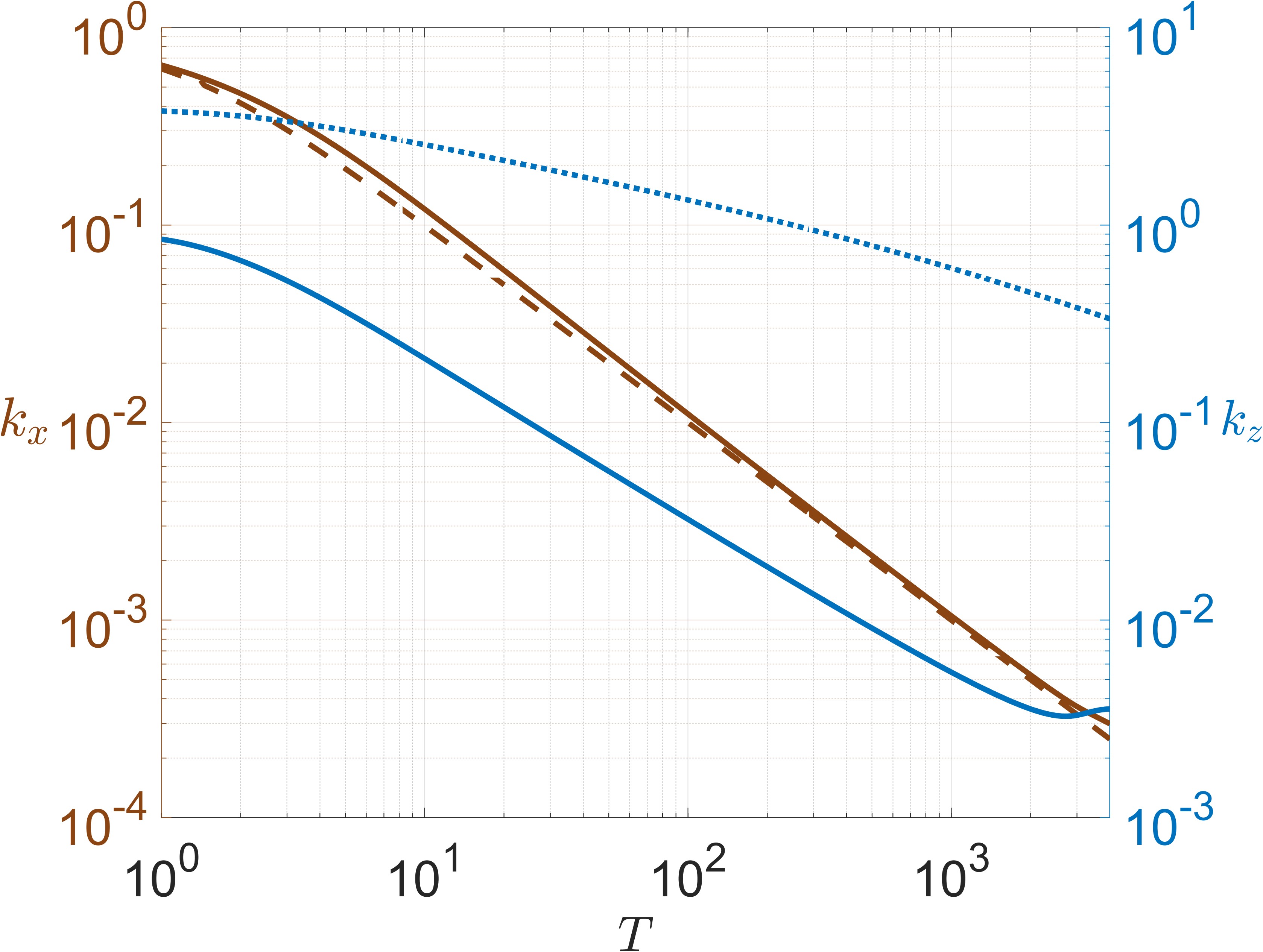}
\caption{Left: optimal growth vs $T$ at $\Rey=1000$. The solid line is the full 3D optimal $\goG(\Rey)=\max_{(k_x,k_z)}G(k_x,k_z;\Rey)$, the dashed line is $\max_{k_x} G(k_x,0;\Rey)$ (Orr) and the dotted line is $\max_{k_z} G(0,k_z;\Rey)$ (lift-up). Right: the corresponding optimal wavenumbers: solid brown/blue for $k_x/k_z$ for the full 3D optimal, dashed brown for $k_x$ in Orr and dotted blue for $k_z$ in lift up. These plots show that the 3D optimal closely follows the Orr result in terms of both $G$ and $k_x$ with assistance from  lift-up except near the maximum over $T$ (the overall optimal gain $\ooG$).  }
\label{G&k_vs_T}
\end{figure}

%
%
\subsection{Fully 3D optimals}

\begin{figure}
\centering
\includegraphics[width=0.55\linewidth]{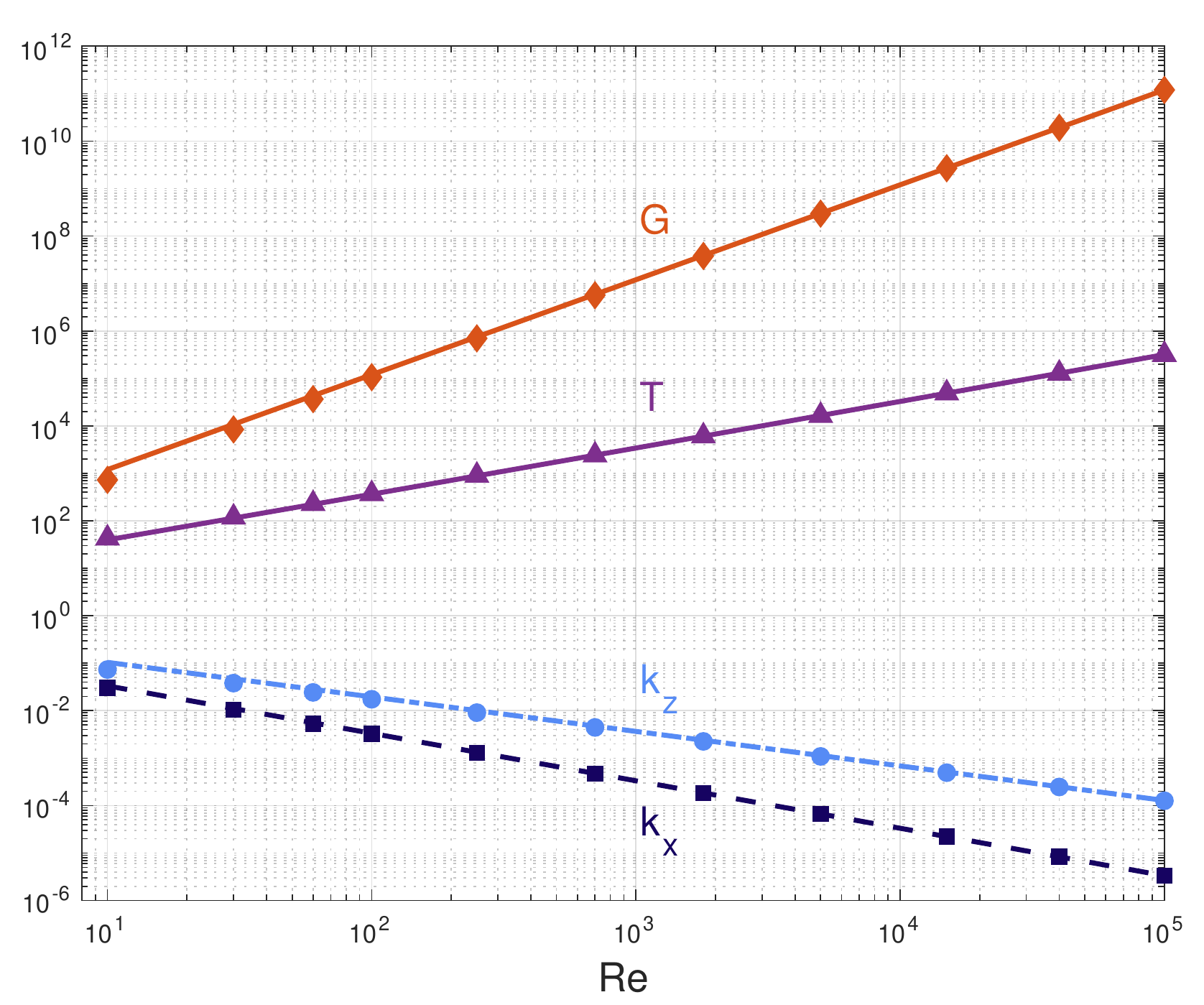}
\caption{This plot shows the overall optimal gain  $\ooG(\beta,\Rey)$ and associated optimising parameters ($k_x, \,k_y$ and $T$) as a function of $\Rey$ for $\beta=0$ in symbols. The asymptotic predictions of (\ref{3D_asym}) are the lines.}
\label{LargeRe}
\end{figure}

Figure \ref{EarlyNoStreakFig} shows the growth landscape over $(k_x,k_z)$ plane for $T=5$ and $\Rey=200$. The optimal $k_x$ wavenumber is close to the (2D) Orr optimal value ($k_z \rightarrow 0$ on the plot) but the optimal $k_z$ is an order of magnitude smaller than the corresponding lift-up value ($k_x \rightarrow 0$). This observation is robust over different $T$ as shown in Figure \ref{G&k_vs_T}(b) and becomes more exaggerated as $T$ increases towards the maximum at $T=O(\Rey)$.
It is also noticeable in \ref{G&k_vs_T}(a) that this overall optimal gain $\ooG$ (maximised over all $k_x,k_z$ and $T$) occurs at about the same time as the Orr maximum (this is clearly not true for lift-up). This and the $k_x$ observation suggests using the large $\Rey$ results  in (\ref{Orr_asym}) for Orr as a starting point for analysing the full 3D optimal. Assuming that the initial shearwise vorticity is zero, this leads to 
\begin{equation}
\ooG^{\text{3D}}= \frac{9\upi^2}{\mathrm{e}^2}\Rey^2, \quad T^{\text{3D}}= 3\Rey + \alpha \Rey^{9/11}, \quad k_x^{\text{3D}}= \frac{1}{3\Rey}, \quad k_z^{\text{3D}}= \gamma \Rey^{-8/11}, \label{3D_asym}
\end{equation}
where $\alpha:=3^{6/11}/\upi^{2/11}\approx 1.4786$ and $\gamma:=\upi^{3/11}/3^{9/11}\approx 0.5562$. These asymptotic predictions match the numerical computations very well even down to $\Rey=O(30)$ - see Figure \ref{LargeRe}, also providing {\it a posteriori} justification for assuming the optimal has $\hat{\eta}(0)=0$. The formulae in (\ref{3D_asym})  give optimal wavenumbers at $\Rey=100$ of  $(k_x,k_z)=(3.33, 19.53) \times 10^{-3}$ which are consistent with figure 1(a) of \cite{Jiao-21}.

%
%

%
%

\section{Early Time Transient Growth in Streaky Flow ($\beta \neq 0$)} \label{streak}

We now shift our focus to streaky unbounded constant shear flow and the transient growth possible over $O(1)$ times as discussed in \S \ref{paramsec}. The required optimals will prove to be fully 3-dimensional but we start by looking at the reduced 2-dimensional (streamwise-independent $k_x=0$) problem which isolates the unfamiliar `push-over' mechanism.

%
%
\subsection{2-Dimensional ($k_x=0$) Push-over Growth} \label{2DTG}

With $k_x=0$, the wavenumbers are constant so Orr is absent and it is simpler to work with the velocity field directly.
The $\hat{v}$ and $\hat{w}$ momentum equations together with  continuity imply $\hat{p}_m=0$ for all $m$, leaving the 3 evolution equations
\begin{align}
\dot{\hat{u}}_m &=  - \frac{k_m^2}{\Rey}\hat{u}_m - \hat{v}_m + 
\frac{\mathrm{i}\beta k_z}{2}(\hat{w}_{m+1}-\hat{w}_{m-1}),       \label{kx0udiff} \\
\dot{\hat{v}}_m &= - \frac{k_m^2}{\Rey}\hat{v}_m,  \qquad
\dot{\hat{w}}_m = - \frac{k_m^2}{\Rey}\hat{w}_m, \label{kx0vwdiff}
\end{align}
where $k^2_m=1+m^2 k_z^2$. These indicate that $\hat{v}_m$ and $\hat{w}_m$ simply viscously decay in time allowing explicit solutions to be written down as follows
\begin{align}
\hat{u}_m = \mathrm{e}^{-k_m^2 t/\Rey}\bigg\{ U_m-V_mt 
+ \frac{\mathrm{i} \Rey \beta}{2k_z} \bigg[ &\frac{W_{m-1}}{1-2m}\bigg(\mathrm{e}^{-(1-2m)k_z^2 t/\Rey}-1 \bigg) \nonumber\\ 
-                                &\frac{W_{m+1}}{1+2m}\biggl(\mathrm{e}^{-(1+2m)k_z^2 t/\Rey}-1 \biggr) \bigg] \bigg\},  \label{usol} \\
\hat{v}_m = V_m \mathrm{e}^{-k_m^2 t/\Rey},& \qquad
\hat{w}_m = W_m \mathrm{e}^{-k_m^2 t/\Rey}, \label{vwsol}
\end{align}
where $U_m:=\hat{u}_m(0)$, $V_m:=\hat{v}_m(0)$ and $W_m:=\hat{w}_m(0)$ are the initial conditions which, by continuity, satisfy $V_m+mk_zW_m=0$.  Taking $\Rey \rightarrow \infty$ gives the simple $\hat{u}$ equation
\begin{gather}
\hat{u}_m = U_m-V_mt + \frac{\mathrm{i}\beta k_z t}{2} \left( W_{m+1} - W_{m-1}\right) .  \label{usolearly1} 
\end{gather}
The push-over effect is produced by the forcing term proportional to the spanwise streak shear, $\beta k_z$, and is very similar to the $-V_m t$ lift-up forcing (normalised to have unit shear) although push-over drives neighbouring spanwise modes while lift-up is direct. 
Lift-up can be turned off by setting $V_m=0$ for $\forall m$ which, through continuity, forces $W_m=0$ for $m \neq 0$.  Setting $U_m=0$ for all $m$ as well appears an extreme choice for initial conditions but  numerical computations indicate that it is close to optimal for $k_x=0$ (e.g. at $\beta=1$, $T=5$ and $\Rey=200$, the error using the extreme choice is only $\approx 1\%$ of the global optimal gain value). When only $W_0$ of the initial conditions is non-zero, the gain is
\begin{equation}
\hat{G}^{\text{push}} = \mathrm{e}^{-2T/\Rey}\left[ 1+\frac{\Rey^2\beta^2}{2k_z^2}\left(1-\mathrm{e}^{-Tk_z^2/\Rey}\right)^2\right] \sim 1+\frac{1}{2}\beta^2 k_z^2 T^2 \quad {\rm as}\, \Rey \,\rightarrow\, \infty.  \label{Push_T_1}
\end{equation}
assuming $T \ll \Rey$ and  $k_z^2 \ll \Rey/T$ for the asymptotic simplification (we use  a $\hat{\quad}$ to indicate a quantity not optimised over all initial conditions).
This mirrors the result for lift-up: when $\beta=0$ is used, choosing only $(U_1,V_1,W_1)=(0,1, -1/k_z)$ to be non-zero gives a gain of 
\begin{equation}
\hat{G}^{\text{lift-up}} \sim 1+k_z^2T^2/(1+k_z^2) 
\label{Lift_up_T_1}
\end{equation}
at large $\Rey$. Hence push-over seems to behave very much like lift-up in both the growth possible and the timescales, i.e. growth is ultimately limited by the viscous decay of the spanwise velocities so push-over  is `slow'.

The overall optimal gain in (\ref{Push_T_1}) can be found by optimising over both $k_z$ and $T$: $\partial \hat{G}^{\text{push}}/\partial k_z = \partial \hat{G}^{\text{push}}/\partial T = 0$ requires
\begin{gather}
2k_z^2Te^{-k_z^2T/\Rey}-\Rey\left(1-\mathrm{e}^{-k_z^2T/\Rey}\right)=0, \label{2DPOopt1} \\
\Rey^2 \beta^2 \left(1-\mathrm{e}^{-k_z^2T/\Rey}\right)\left[\mathrm{e}^{-k_z^2T/\Rey}(1+k_z^2)-1\right] = 2k_z^2. \label{2DPOopt2}
\end{gather}
The first of these is solved by noticing that the equation reduces to $2\zeta + 1 - \mathrm{e}^{\zeta}=0$ with 
$\zeta:=k_z^2T/\Rey$
which has the  unique positive solution $\zeta \approx 1.256$. 
%
Then, using $\mathrm{e}^{-\zeta}=(1+2\zeta)^{-1}$ in (\ref{2DPOopt2}) gives the optimal wavenumber as $\hat{k}^{push}_z=\sqrt{2\zeta}$, and thus the maximising time and maximal gain for only-non-vanishing-$W_0$ initial conditions as
\beq
\hat{T}^{push}=\tfrac{1}{2}\Rey \qquad \text{and} \qquad \hat{\ooG}^{push}=\frac{\zeta \beta^2}{\mathrm{e}(1+2\zeta)^2} \Rey^2.
\label{2dPO3}
\eeq
This again shows that the push-over mechanism, in the absence of Orr ($k_x=0$) and with initial conditions chosen to eliminate lift-up, possesses the same scalings as the lift-up mechanism.

%
%
\begin{figure}
\centering
    \begin{subfigure}{0.85\textwidth}
        \centering
        \includegraphics[width=\textwidth]{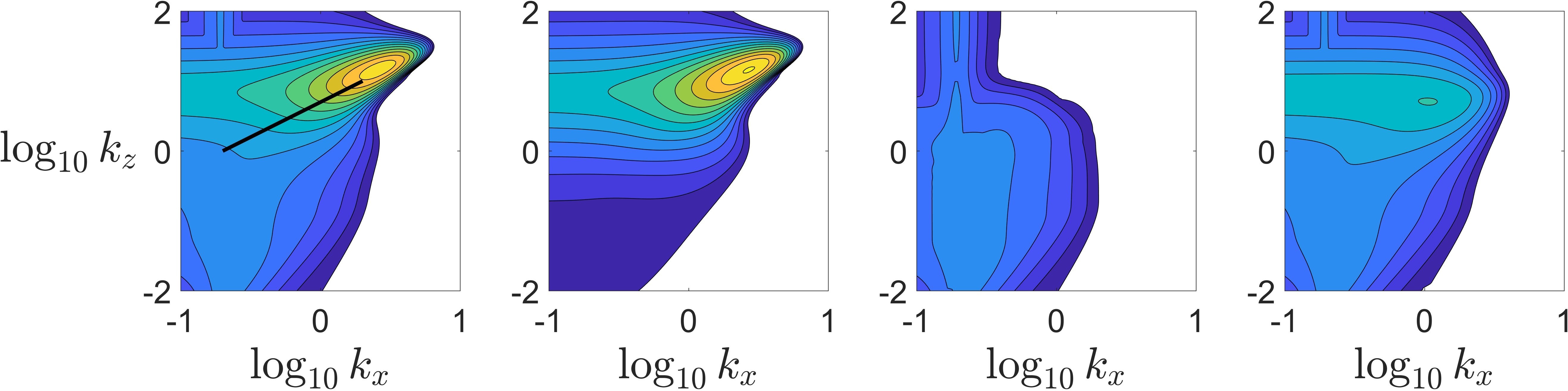}
    \end{subfigure}
    \begin{subfigure}{0.104\textwidth}
        \centering
        \includegraphics[width=\textwidth,trim=0cm 1.3cm 0.05cm 0cm, clip]{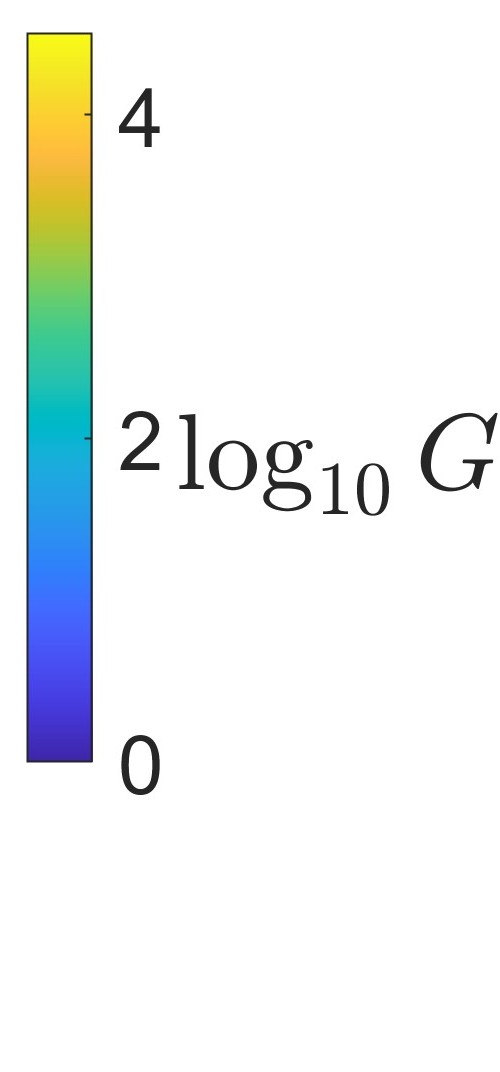}
    \end{subfigure}
\caption{Contour plots of the optimal gain over the $(k_x,k_z)$ wavenumber plane for the parameter set $\beta=1$, $T=5$ and $\Rey=200$. The truncation value is set to $M=20$ and the contour levels are the same across all the plots.  Plot (a) shows the full problem (all terms included), while plots (b), (c) and (d) have the lift-up, push-over and Orr mechanisms removed, respectively. The black line traces the wavenumber ranges used for figure \ref{ConPhys}. The point of this  figure is to show that the large growth observed is substantially suppressed when either the push-over or Orr mechanisms are removed, but is largely independent of the lift-up mechanism. }
\label{comp4}
\end{figure}

%
%
\subsection{Transient Growth of Three-Dimensional Perturbations} \label{3DTG}

\begin{figure}
\centering
\includegraphics[width=0.9\textwidth]{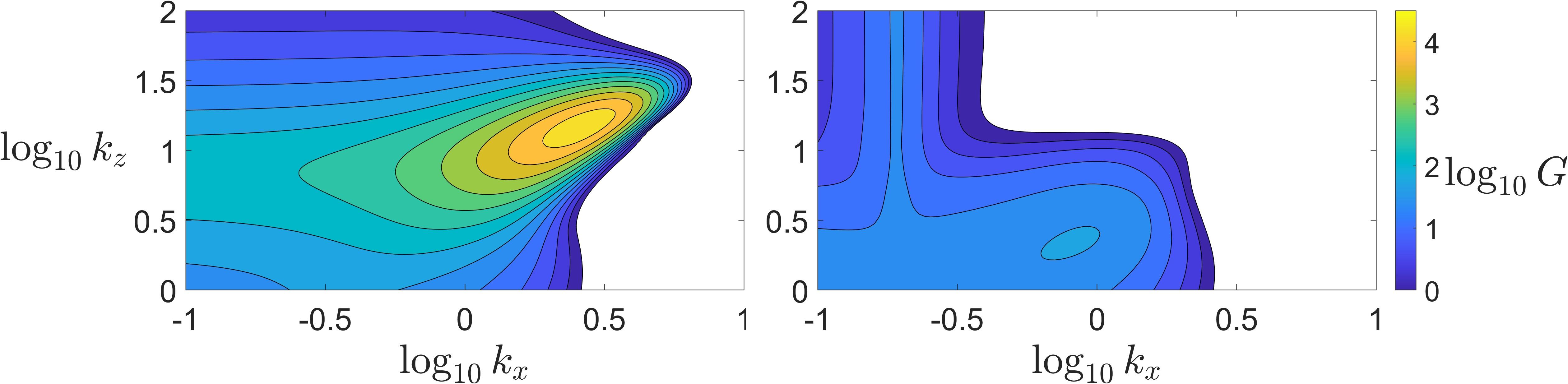}
\caption{Contour plots of the optimal gain over the $(k_x,k_z)$ wavenumber plane for $\beta=1$, $T=5$ and $\Rey=200$. Both use $M=20$ in the model, but the first plots the optimal `sinuous' disturbances, while the second plots the optimal `varicose' disturbances (recall that this terminology refers to the symmetry or antisymmetry of $u$ about $z=0$). This plot demonstrates that all of the large growth occurs for sinuous perturbations.}
\label{VariSinu}
\end{figure}

Larger energy growth occurs for three-dimensional disturbances, where all three mechanisms (lift-up, push-over and Orr) can play a role. Figure \ref{comp4} displays contours of the optimal gain in wavenumber space for four different scenarios at representative parameter values ($\beta=1$, $T=5$ and $\Rey=200$).
The left-most plot shows the gain at $T$ at a given pair of wavenumbers $(k_x,k_z)$ maximised over all initial conditions for the full equations. The other three plots extending to the right have the three growth mechanisms excluded in turn by artificially removing the terms which drive them: figure \ref{comp4}(b) has lift-up suppressed by removing $\partial {\boldsymbol U_B}/\partial y$ from the equations;  figure \ref{comp4}(c) has push-over suppressed by removing $\partial {\boldsymbol U_B}/\partial z$; and figure \ref{comp4}(d) has the streak advection (\,the $\beta \cos (k_z z)\partial {{\boldsymbol u}}/\partial x$ term\,) removed to suppress the `spanwise' Orr mechanism (removing all of the advection changes the whole character of the flow). All optimal disturbances here are sinuous  with varicose disturbances always producing less growth in this augmented Kelvin model across the parameters considered: e.g. see figure \ref{VariSinu}.

%
%

%
%
\begin{figure}
\centering
\includegraphics[width=0.65\linewidth]{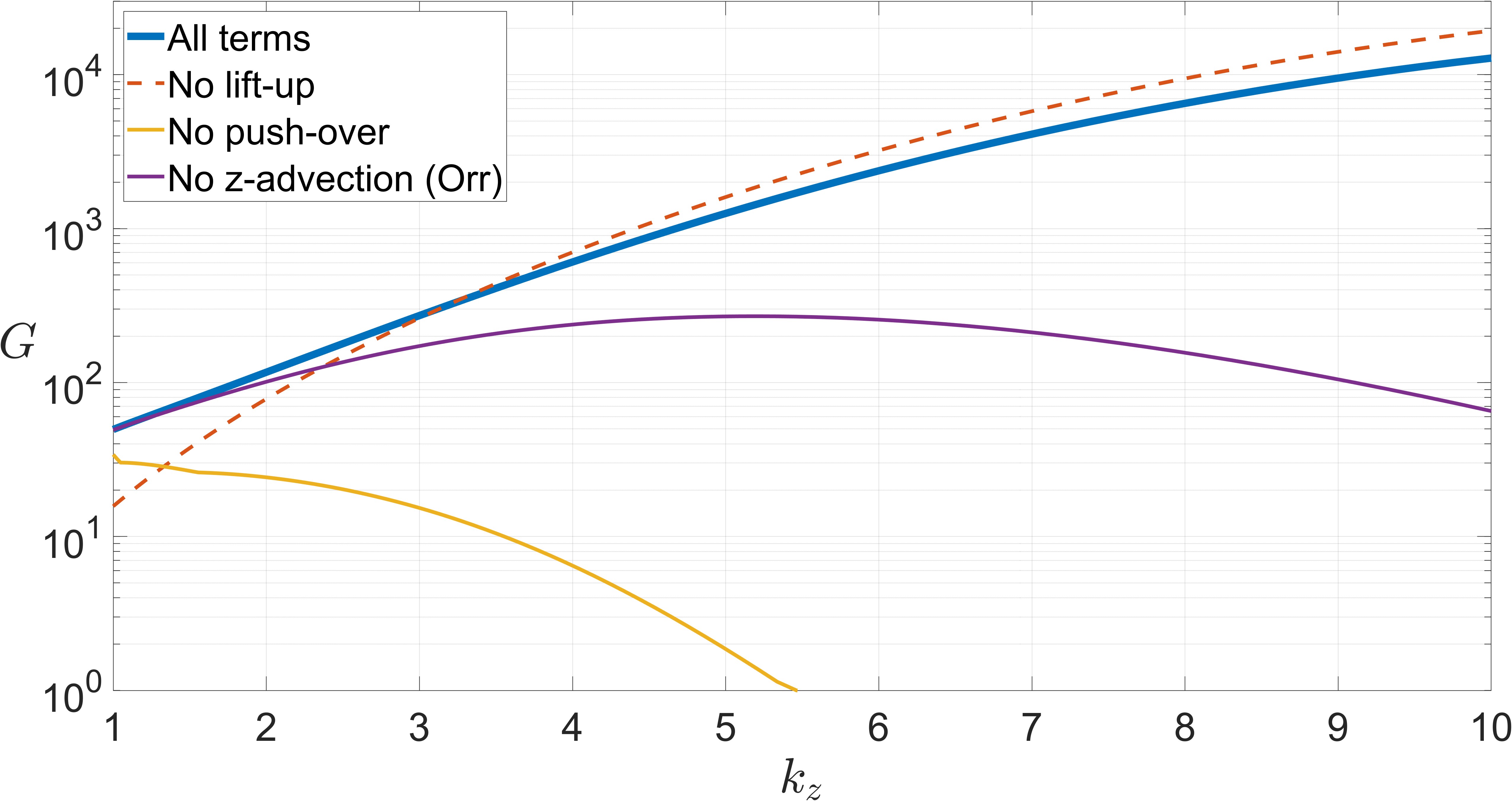}  
\caption{A plot showing the optimal gain as $k_z$ is varied, with $k_x=k_z/5$ and parameters $\beta=1$, $T=5$ and $\Rey=200$. The range of $k_z$ used here is indicated in figure \ref{comp4}(a) using a black line. The blue line represents the gain when all terms are included in the equations (corresponding the the contours in figure \ref{comp4}(a) ). The red line shows the gain when the lift-up mechanism is removed, while the yellow line shows gain when the push-over mechanism is removed and the purple line shows gain when the spanwise Orr mechanism is removed (which correspond to the contours in figure \ref{comp4}(b), \ref{comp4}(c) and \ref{comp4}(d), respectively). This plot demonstrates that both the push-over and spanwise Orr mechanisms are crucial for large growth of kinetic energy, even down to some moderate wavenumbers. It also shows that for smaller wavenumbers, a switch occurs and lift-up becomes the dominant growth process.}
\label{ConPhys}
\end{figure}

A key feature of the figure \ref{comp4}(a) is the global optimal gain indicated by the yellow region, which reaches values of $\goG \approx 10^4$ at $k_x  \approx 2.7 $ and $k_z\approx 15$. A comparison of this peak and the surrounding region to the same location in the other plots reveals that removing the lift-up mechanism leaves the optimal gain contours almost completely unaltered, while removing either the push-over or  spanwise Orr mechanisms reduces the size of gain significantly. This importance of push-over and lack of importance of lift-up for a streaky base flow is fully consistent with  earlier observations \citep{Loz-21,Mar-24}.

The optimal wavenumbers for this global optimal are, however,  on the large side for the lower part of the log-layer (see section \ref{paramsec}) but a line drawn along the major axis of the contours - approximately $k_z=5k_x$ - suggests a direction of `shallowest' descent in the wavenumber plane (shown as a black line in figure \ref{comp4}a for $1<k_z<10$).  Figure \ref{ConPhys} shows how the optimal gain behaves  along this line for $1<k_z<10$ when the various mechanisms are excluded.  At $k_z=1$, removing lift-up is actually more detrimental to the gain than removing either spanwise Orr or push-over, but for $k_z \gtrsim 3$, the situation is reversed and mimics that for the global optimal gain. The figure also shows that the presence of lift-up now actually hinders growth (the crossing of blue and red lines at $k_z \approx 3$). This effect is also present in figure \ref{comp4}(b) and is consistent with the observations of \cite{Loz-21} and figure 7 in \cite{Mar-24}. 

So, the conclusion is, in this unbounded streaky shear flow at these parameters, wavenumber pairings reaching down to those appropriate for buffer-layer dynamics show the clear symbiosis of Orr and push-over mechanisms and the unimportance of lift-up as highlighted by the global optimal gain pairing.

%
%
\subsection{A 2-Variable Model of the Push-over \& Orr Interaction} \label{simpmodsec}

Remarkably, it is possible to strip the model down to just 2 fields and still retain  the symbiotic interaction between push-over and Orr.  There are 3 distinct steps which achieve this. Firstly, the spanwise truncation $M$ is reduced down to $1$ decreasing the number of (complex-valued) fields from $2(2M+1)$ (with $M=20$ or so) down to 6. Secondly, the symmetry: $\hat{v}_0=0$, $\hat{v}_1=-\hat{v}_{-1}$ and $\hat{\eta}_1=\hat{\eta}_{-1}$ appropriate for the sinuous mode \citep[discussed in ][]{Sch-02} is imposed so 6 becomes 3 complex fields. These evolve as follows
%
%
%
\begin{align}
%
%
\frac{d\hat{v}_1}{dt} &= -\frac{k_1^2}{\Rey}\hat{v}_1 
+\frac{2k_x(1-k_xt)}{k_1^2}\hat{v}_1
-\frac{\beta k_z (1-k_xt)}{k_1^2}\hat{\eta}_0,  \label{dv1} \\
%
%
\frac{d\hat{\eta}_0}{dt} &= -\frac{k_0^2}{\Rey}\hat{\eta}_0 
-\overbrace{\mathrm{i}\beta k_x \hat{\eta}_1}^{{\color{red} \text{Orr}}}
+\overbrace{\frac{\beta k_x^2 k_z(1-k_xt)}{h_1^2}\hat{v}_1}^{{\color{red}\text{Orr}}}
+\overbrace{\frac{\mathrm{i}\beta k_x k_z^2}{h_1^2}\hat{\eta}_1}^{{\color{red}\text{Orr}}}, \label{deta0} \\
%
%
\frac{d\hat{\eta}_1}{dt} &= -\frac{k_1^2}{\Rey}\hat{\eta}_1  
- \underbrace{\frac{1}{2}\mathrm{i}\beta k_x \hat{\eta}_0}_{{\color{red}\text{Orr}}}
- \underbrace{\mathrm{i} k_z \hat{v}_1}_{\text{lift up}}
+ \underbrace{\frac{\mathrm{i}\beta k_z^2}{2k_x}\hat{\eta}_0}_{{\color{blue}\text{push over}}}, \label{deta1} 
\end{align}
where, recall, $k_0^2:=k_x^2+(1-k_xt)^2$,  $k_1^2:=k_0^2+k_z^2$ and $h_1^2:=k_x^2+k_z^2$. Here the Orr, lift-up and push-over terms are labelled in the last two equations and the Orr terms only rely on the streak flow (indicated by being premultiplied by $\beta$). 
The fact that the term involving $\hat{v}_1$ in (\ref{deta0}) is only an Orr term, and that (\ref{deta0}) contains no push-over contributions is not obvious; see the Appendix  for details.

%
\begin{figure}
\centering
    \begin{subfigure}{0.85\textwidth}
        \centering
        \includegraphics[width=\textwidth]{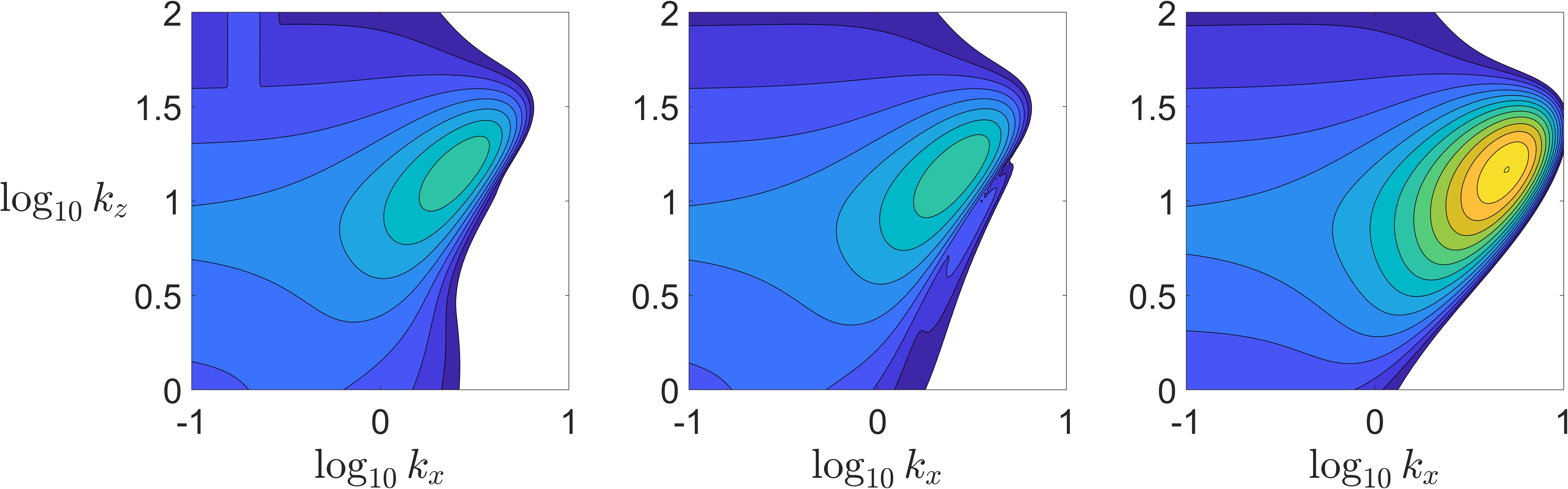}
    \end{subfigure}
    \begin{subfigure}{0.107\textwidth}
        \centering
        \includegraphics[width=\textwidth,trim=0 1.5cm 0cm 0cm, clip]{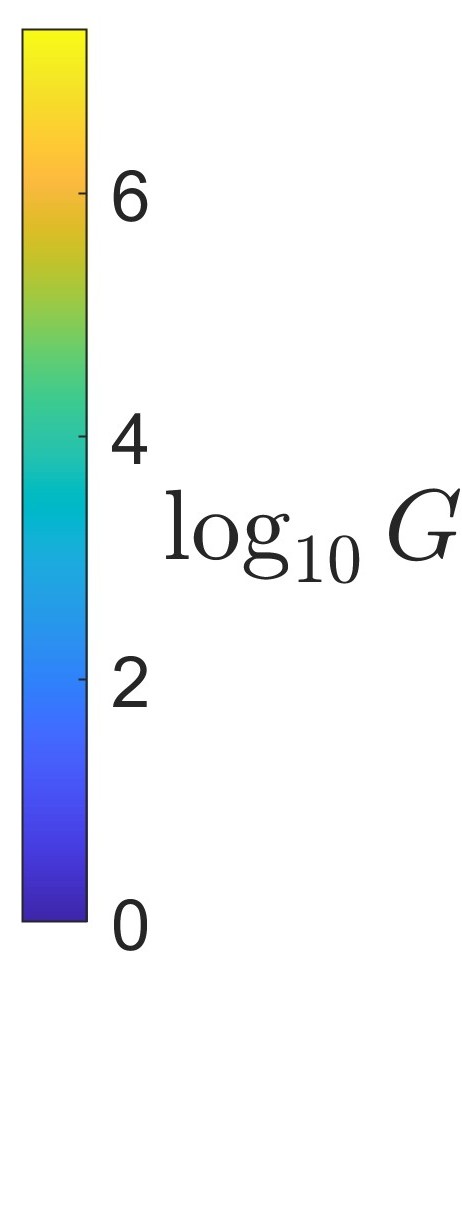}
    \end{subfigure}
    \caption{A sequence of contour plots showing the optimal gain in wavenumber space, close to the optimal pair, for the parameter set $\beta=1$, $T=5$ and $\Rey=200$. The colour bar is shown on the right hand side and is the same for all three plots. Plot (a) has truncation value $M=20$, (b) has $M=1$ with the sinuous symmetry used and (c) uses the further simplification of having $\hat{v}_1$ removed. This figure demonstrates that the underlying growth process is present in all three cases with the reduced model harbouring significantly enhanced growth.}
\label{comp3}
\end{figure}

\begin{figure}
\centering
    \begin{subfigure}{\textwidth}
        \includegraphics[width=0.95\textwidth]{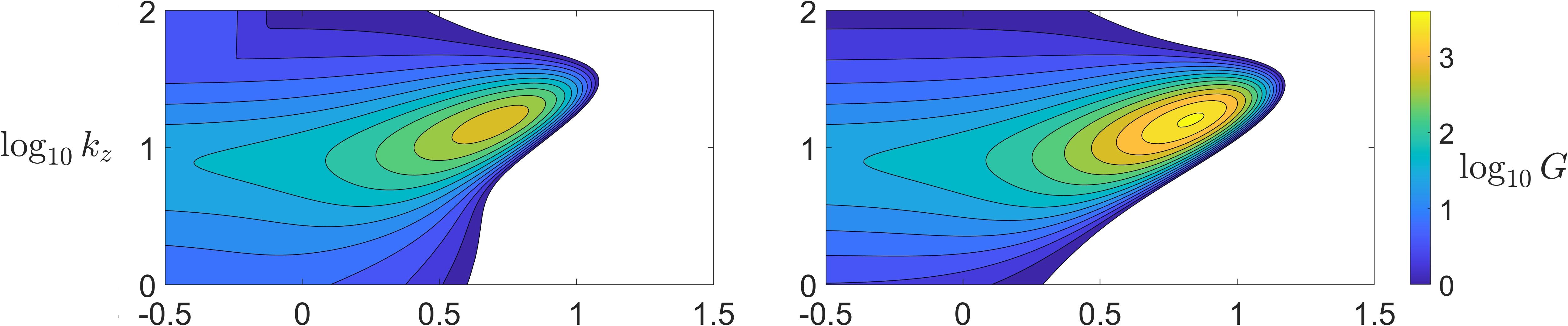}
    \end{subfigure}
    \hfill
    \vspace{0.1cm}
    \begin{subfigure}{\textwidth}
        \includegraphics[width=0.956\textwidth]{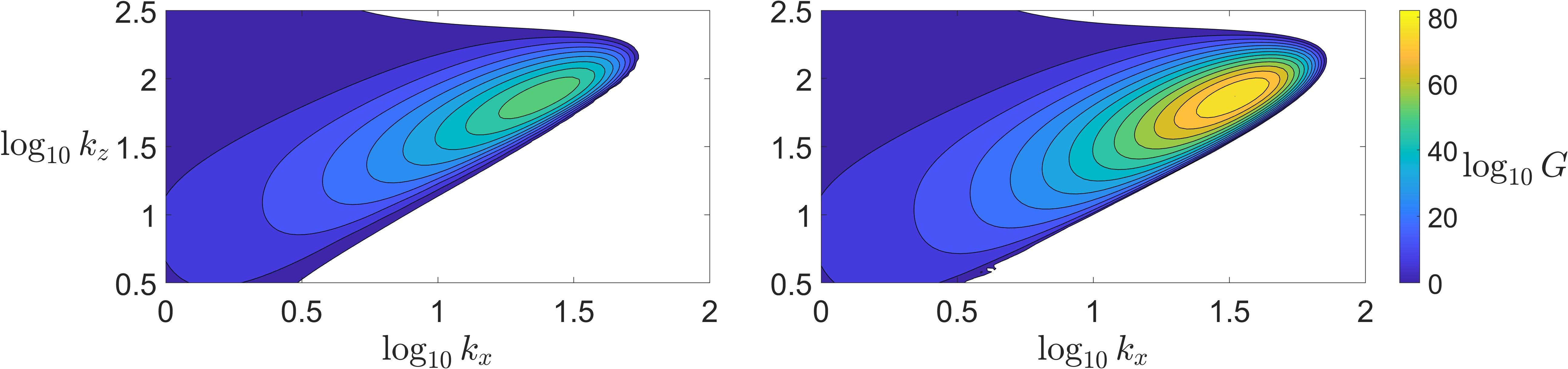}
    \end{subfigure}
\caption{Contour plots in wavenumber space of the optimal gain for $\beta=1$(upper) and $\beta=5$ (lower), with other parameters set to $T=2$, $\Rey=100$. The first column shows the full model with truncation $M=20$ while the second column shows the reduced two-variable model given by equations (\ref{2vardiff_a}) and (\ref{2vardiff_b}). }
\label{VaryBeta}
\end{figure}

Finally, $\hat{v}_1$ can be jettisoned meaning \eq{dv1} is ignored and $\hat{v}_1$ set to zero in the $\hat{\eta}$ equations, leaving simply
\begin{align}
\frac{d\hat{\eta}_0}{dt} &=  -\frac{k_0^2}{\Rey}\hat{\eta}_0 
\,-\,
\overbrace{ \frac{\mathrm{i} \beta k_x^3}{k_x^2+k_z^2} }^{{\color{red}\text{Orr}}}
 \hat{\eta}_1, \label{2vardiff_a}\\
\frac{d\hat{\eta}_1}{dt} & =  -\frac{k_1^2}{\Rey}\hat{\eta}_1 
\,+\, \frac{1}{2}\mathrm{i} \beta k_x\bigg[
\underbrace{\frac{k_z^2}{k_x^2}}_{{\color{blue}\text{push over}}}\,
\underbrace{-1}_{{\color{red}\text{Orr}}}\bigg]\hat{\eta}_0. \label{2vardiff_b}
\end{align}
The {\em a posteriori} justification for such a drastic reduction is provided by figure \ref{comp3} which shows the gain in the full ($M=20$) system (the same data as figure \ref{comp4}(a) but replotted over a smaller wavenumber area and with different contour levels), the gain produced by the three-variable model (\ref{dv1})-(\ref{deta1}), and the reduced model (\ref{2vardiff_a})-(\ref{2vardiff_b}). This comparison clearly shows that: i) the $M=1$ reduction is surprisingly effective in retaining the gain behaviour 
and ii) the reduced model also retains the global optimal gain feature and in fact, as also observed,  experiences significantly enhanced growth with  lift-up removed.  
Figure \ref{VaryBeta} shows this picture is repeated for smaller $\Rey=100$ and target time of $T=2$ but a larger $\beta=5$ value (note the increase in contour scale for $\beta=5$ compared with $\beta=1$ which is explained by the scaling in  (\ref{largegrowth})\,).

%
%

For $Tk_1^2 \ll \Rey$, diffusion can be dropped in (\ref{2vardiff_a})-(\ref{2vardiff_b}) and then setting $\hat{\eta}_0=\eta_0$ and $\hat{\eta}_1=i \eta_1$ for convenience gives just
\begin{equation}
\frac{d}{dt}\left(
\begin{array}{c}
\eta_0\\
\eta_1
\end{array}
\right) =
\left( \begin{array}{cc}
0  & \beta k_x^3/(k_x^2+k_z^2) \\
\tfrac{1}{2}\beta (k_z^2-k_x^2)/k_x & 0 \\
\end{array} \right)
\left(
\begin{array}{c}
\eta_0\\
\eta_1
\end{array}
\right).
\label{2varnodiff}
\end{equation}
To arrange for the Euclidean norm to correspond to the energy, the vorticities need to be rescaled - $\tilde{\eta}_0:=\eta_0/h_0$ and $\tilde{\eta}_1:=\sqrt{2} \eta_1/h_1$ - so that 
\begin{equation}
E:=\tfrac{1}{2}(|\hat{\eta}_0/h_0|^2+2|\hat{\eta}_1/h_1|^2)=\tfrac{1}{2}(|\tilde{\eta}_0|^2+|\tilde{\eta}_1|^2)
\end{equation}
and the system becomes
\begin{equation}
\frac{d}{dt}\left(
\begin{array}{c}
\tilde{\eta}_0\\
\tilde{\eta}_1
\end{array}
\right)  = \LL \left(
\begin{array}{c}
\tilde{\eta}_0\\
\tilde{\eta}_1
\end{array}
\right)
:=
\left( \begin{array}{cc}
0  & a \\
b & 0 \\
\end{array} \right)
\left(
\begin{array}{c}
\tilde{\eta}_0\\
\tilde{\eta}_1
\end{array}
\right).
\label{re2varnodiff}
\end{equation}
where 
\begin{equation}
a:= \frac{\beta k_x^3}{k_x^2+k_z^2}\frac{h_1}{h_0\sqrt{2}}=\frac{\beta k_x^2}{\sqrt{2(k_x^2+k_z^2)}}
, \qquad
b:= \frac{\beta (k_z^2-k_x^2)}{2 k_x}\frac{h_0\sqrt{2}}{h_1}=\frac{ \beta(k_z^2-k_x^2)}{ \sqrt{2(k_x^2+k_z^2)}}
\end{equation}
This system can produce energy growth in 2 distinct and independent ways.\\
\begin{enumerate}
\item Algebraic growth due to the non-normality of $\LL$. This occurs when the magnitudes of the off diagonal elements of $\LL$ do not match i.e.
\begin{equation}
\frac{a^2}{b^2} =\left(\frac{k_x^2}{k_z^2-k_x^2}  \right)^2\neq 1 \quad \Rightarrow \quad k_z \neq \sqrt{2} k_x
\end{equation}
which ensures $\LL \LL^T \neq \LL^T \LL$ - the defining property that $\LL$ is a non-normal matrix.\\
\item Exponential growth due to a linear instability with growth rate $\sigma$ where
\begin{equation}
\sigma^2:= ab = \frac{1}{2} \beta^2 k_x^2 \left( \frac{k_z^2-k_x^2}{k_z^2+k_x^2} \right).
\end{equation}
This needs $k_z > k_x$ and so push over dominates streak Orr in (\ref{2vardiff_b}). Since only streak Orr operates in (\ref{2vardiff_a}), this clearly shows that push over and streak Orr are needed for this linear instability.\\
\end{enumerate}

%
%
\subsection{2-Variable Energy Growth} \label{4.4}

Either mechanism can act in isolation or together depending on the wavenumber pair $(k_x,k_z)$. In the model studied here, the wavenumbers $k_x$ and $k_z$ are naturally chosen so that both mechanisms are operative as this creates the largest growth. In particular, in the example plotted in Figure \ref{comp3} - $(Re,T,\beta)=(200,5,1)$ - the optimal wavenumbers are $(k_x,k_z) \approx (2.7, 15)$ for the M=20 and M=1 models so that  $a \approx 0.34$ and $b \approx 10.1$.
Hence $k_z > k_x$ and there is actually a significant difference in magnitude between the off-diagonal elements. 
To characterise how much energy growth this produces, the 2 differential equations in (\ref{re2varnodiff}) can be straightforwardly integrated to give
\begin{equation}
\left(
\begin{array}{c}
\tilde{\eta}_0(t)\\
\tilde{\eta}_1(t)
\end{array}
\right)  = \A \left(
\begin{array}{c}
\tilde{\eta}_0(0)\\
\tilde{\eta}_1(0)
\end{array}
\right)
:=
\left( \begin{array}{cc}
\cosh \sigma t  & \sqrt{a/b} \sinh \sigma t \\
\sqrt{b/a} \sinh \sigma t & \cosh \sigma t \\
\end{array} \right)
\left(
\begin{array}{c}
\tilde{\eta}_0(0)\\
\tilde{\eta}_1(0)
\end{array}
\right)
\label{phi}
\end{equation}
where $\A=e^{\LL t}$. The maximum growth possible, $\oG(T)$, at $t=T$ for given wavenumbers $(k_x,k_z)$ is then the largest eigenvalue of the real symmetric matrix $\A^T(T) \A(T)$ (here $\A^T$ indicates the transpose of $\A$ rather than anything to do with the target time $T$). This is just the larger of the two real positive eigenvalues of the quadratic
\begin{equation}
\lambda^2-\left[2 \cosh^2 \sigma T+ \left( \frac{a}{b}+\frac{b}{a}\right) \sinh^2 \sigma T \right]\lambda +1 =0.
\end{equation}
If $\lambda_{max}>1$ then $\lambda_{min}=1/\lambda_{max}<1$, that is, there is only one growth mechanism for a given $(k_x,k_z)$ pairing and no growth implies $\lambda_{max}=\lambda_{min}=1$.  Assuming either large $\sigma T$ when $\sigma^2=ab>0$ or  $|a/b| \neq O(1)$ for $0 > \sigma^2=-\omega^2$,
\begin{equation}
\oG(T,k_x,k_z;\beta):=\lambda_{\rm max} 
\approx 
2 \cosh^2 \sigma T+ \left( \frac{a}{b}+\frac{b}{a}\right) \sinh^2 \sigma T.
\label{G_max}
\end{equation}
This shows either linear instability enhanced by initial transient growth as $a/b+b/a \geq 2$ 
or only transient growth in a linearly stable system (ignoring the singular case of $\sin \omega T=0$). The corresponding optimal condition, which is the associated left eigenvector of $\A^T(T) \A(T)$, depends crucially on whether $a/b \ll 1$ or $\gg 1$. Taking the former and setting $\eps:= \sqrt{a/b} \ll 1$ ($k_z/k_x \gg 1$), the optimal initial condition is 
\begin{equation}
\left(
\begin{array}{c}
\tilde{\eta}_0(0)\\
\tilde{\eta}_1(0)
\end{array}
\right)=
\left(
\begin{array}{c}
1\\
\eps\,{\rm coth} \,\sigma T
\end{array}
\right)
 \quad
 \Rightarrow
 \quad
 \left(
\begin{array}{c}
\tilde{\eta}_0(T)\\
\tilde{\eta}_1(T)
\end{array}
\right)=
\left(
\begin{array}{c}
\cosh \sigma T\\
\frac{1}{\eps}\sinh \sigma T
\end{array}
\right)
\label{TG1}
\end{equation}
along with what it evolves into a time $T$ later (all to leading order in $\eps$). This indicates that energy growth through the non-normality of $\LL$ is produced by initial $\hat{\eta}_0$ generating $\hat{\eta}_1$ as described by equation (\ref{2vardiff_b}). This is consistent with what \cite{Sch-02} found - the initial condition they chose was purely $\hat{\eta}_0$ - but they focussed on the streamwise vorticity as a measure of the growth rather than the shearwise vorticity.

If instead, $a/b \gg 1$ so now $\eps:=\sqrt{b/a} \ll 1$ (or $k_z^2/k_x^2-1 \ll 1$), the optimal initial condition is
\begin{equation}
\left(
\begin{array}{c}
\tilde{\eta}_0(0)\\
\tilde{\eta}_1(0)
\end{array}
\right)=
\left(
\begin{array}{c}
\eps\,{\rm coth} \,\sigma T\\
1
\end{array}
\right)
 \quad
 \Rightarrow
 \quad
 \left(
\begin{array}{c}
\tilde{\eta}_0(T)\\
\tilde{\eta}_1(T)
\end{array}
\right)=
\left(
\begin{array}{c}
\frac{1}{\eps}\sinh \sigma T\\
\cosh \sigma T
\end{array}
\right)
\label{TG2}
\end{equation}
This indicates that energy growth through the non-normality of $\LL$ is now produced by initial $\hat{\eta}_1$ generating $\hat{\eta}_0$ as described by equation (\ref{2vardiff_a}).

%
%
\begin{figure}
\centering
\includegraphics[width=0.85\linewidth]{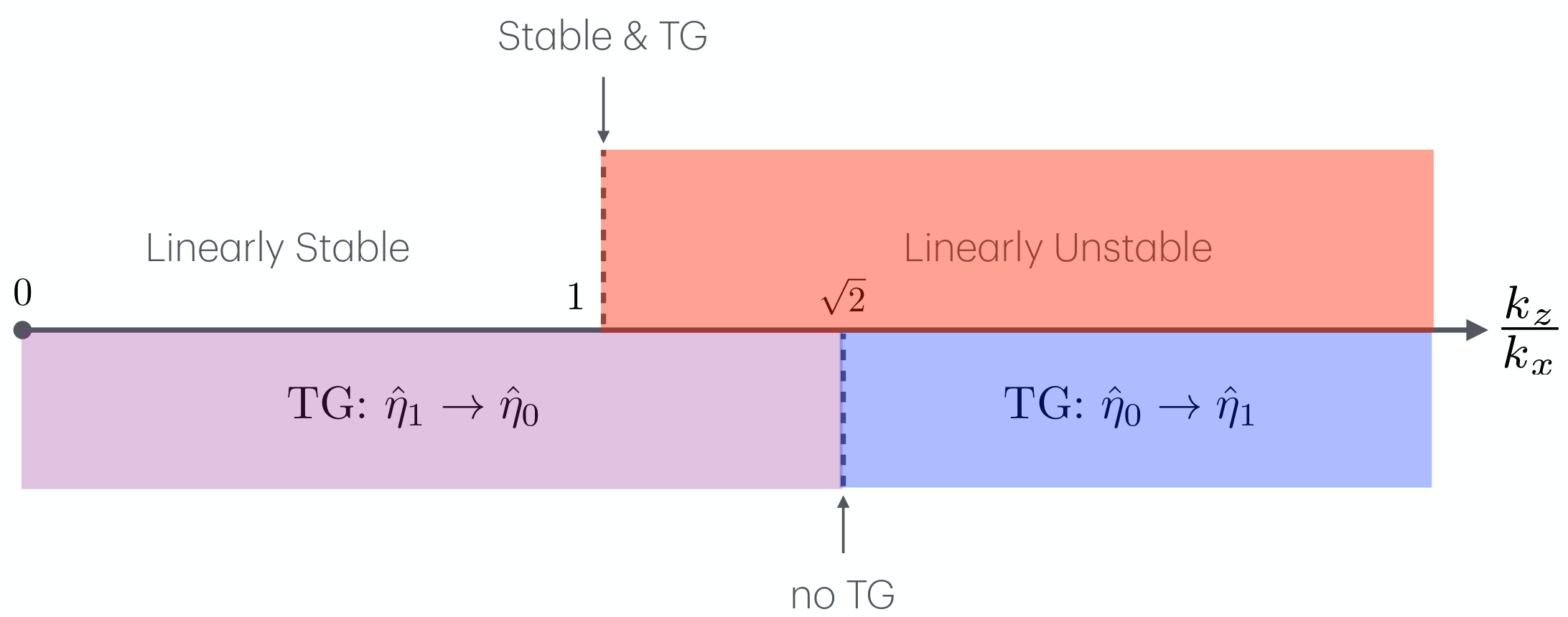}
\caption{The various transient growth (TG) and stability regimes over $0\leq k_z/k_x $. They are  i) $0 \leq k_z/k_x < 1$ - linearly stable and  $\hat{\eta}_1 \rightarrow \hat{\eta}_0$ transient growth; ii) $k_z/k_x = 1$ - linearly stable but unlimited algebraic growth  in $\hat{\eta}_0$; iii) $1< k_z/k_x < \sqrt{2}$ - linearly unstable and  $\hat{\eta}_1 \rightarrow \hat{\eta}_0$ transient growth; iv) $k_z/k_x=\sqrt{2} $ - linearly unstable with no transient growth; and v) $\sqrt{2}< k_z/k_x$ -  linearly unstable and $\hat{\eta}_0\rightarrow \hat{\eta}_1$ transient growth. Sample evolutions are shown for $k_z/k_x=5.6$ in Figure \ref{evolution} and $k_z/k_x=0.95$ in Figure \ref{evolution1} for $(Re,T,\beta)=(200,5,1)$.
}
\label{TG-LI}
\end{figure}

\begin{figure}
\centering
\includegraphics[width=0.95\linewidth]{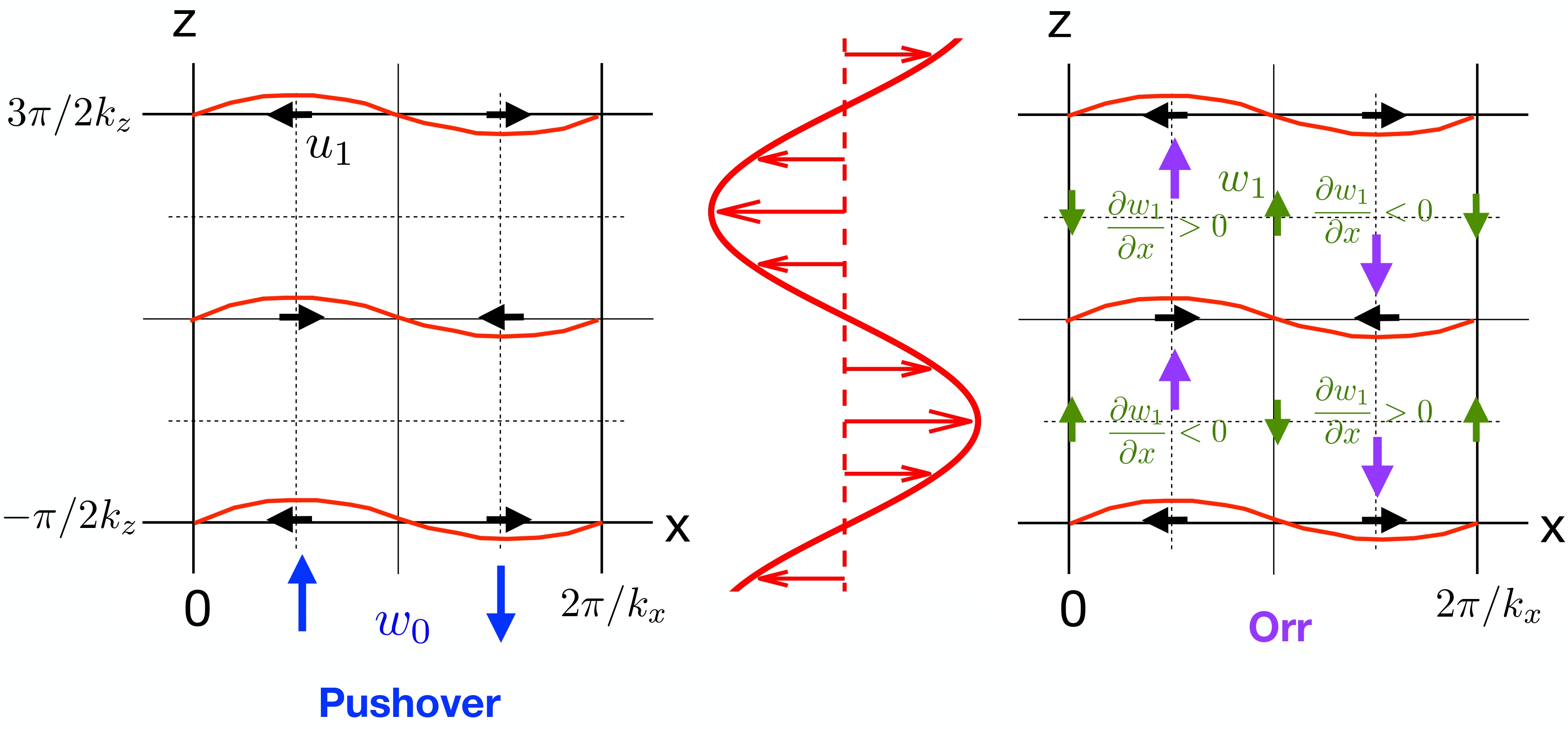}
\caption{Left: $\hat{\eta}_0$ corresponds to a spanwise velocity $w_0 \propto \sin k_x x$ which `pushes over' the streak velocity field $\beta \cos k_z z \,{\bf \hat{x}}$ to produce wavy streaks as shown created by the streamwise velocity anomalies $u_1 \propto \sin k_x x \sin k_z z$ (black arrows). Right: the spatial gradients in the $u_1$ field imply a doubly-periodic pressure field which drives a concomitant spanwise velocity $w_1 \propto \cos k_x x \cos k_z z$ field (green arrows). The advection of this $w_1$ field by the streak velocity - the streak-Orr effect - generates further spanwise velocity (purple arrows) via  the term $-\beta \cos k_z z \partial w_1/\partial x$ which feedbacks positively on $\hat{\eta}_0$ completing the loop. }
\label{mechanism}
\end{figure}

Given instability exists  for $ab>0$ or $k_z > k_x$, there are 5 different scenarios as $k_z/k_x$ varies: see Figure \ref{TG-LI}.  Optimising energy growth over all wavenumbers should select either $1< k_z/k_x < \sqrt{2}$ or $\sqrt{2} < k_z/k_x$ since both are linearly unstable but have different transient growth mechanisms. Computations here suggest it is the latter which contains the optimum which is consistent with where the instability growth rate is larger for fixed $k_x$.
In this case, the transient growth which occurs is described by equation \ref{2vardiff_b} and push over is  the dominant process. This transient growth situation is illustrated in Figure \ref{mechanism} (left). Here, the spanwise velocity perturbation $\hat{w}_0:=\mathrm{i} \hat{\eta}_0/k_x $ `pushes over' (advects) the base streak velocity $\beta \cos (k_z z) \,\hat{{\bf x}}$ to create wavy streaks $\hat{u}_1$ (and so $\hat{\eta}_1$) via the projection of the term $\hat{w}_0\tfrac{\partial}{\partial z} U_B$ onto the $\hat{u}_1$ equation. The (streak) Orr achieves the same effect of generating streamwise- and spanwise-dependent flow from only streamwise-dependent flow - albeit with the opposite and wrong phase for subsequent instability.

%
%

%
%
\subsection{2-Variable Linear Instability} \label{4.5}

The instability looks to be exactly the 2D linear instability of a streak field as modelled by 2D Kolmogorov flow \citep{Arnold60, Meshalkin61}. In its simplest form, Kolmogorov flow consists of a steady forcing,  $\sin \ell y \,{\bf \hat{x}}$ for some integer $\ell$,  applied to a flow over a 2-torus $[0,2\pi/{\hat\alpha}] \times [0,2\pi]$. The corresponding 1-dimensional base flow is exactly the streak flow studied here with the base shear removed and is linearly unstable at high enough $Re$ provided $\hat{\alpha}< \ell$ which is exactly the condition $k_x < k_z$ for instability found in (\ref{2varnodiff}) (e.g. \cite{Marchioro86} and figure 2 in \cite{CK13}). 

The instability is produced by a further mechanism represented by (\ref{2vardiff_a}) which generates $\hat{\eta}_0$ from $\hat{\eta}_1$ to complement the transient growth mechanism. As discussed above, the latter produces wavy streaks which drive a concomitant spanwise field $\hat{w}_1$ through continuity ($k_x \neq 0$): see Figure \ref{mechanism} (right). This spanwise field  $\hat{w}_1$ then drives $\hat{w}_0$ to close the loop by the streak Orr mechanism. that is, streak advection of $\hat{w}_1$ given by the projection of the $\beta \cos (k_z z)\mathrm{i}k_x\hat{w}_1$ term onto the $\hat{w}_0$ equation: again see Figure \ref{mechanism} (right). In this process, the streamwise flow component of the instability, $\hat{u}_1$, is largest at the inflexion points of the streak field where the spanwise shear is maximal, while the spanwise flow component is largest when the streak is largest.  \\

\color{black}

%
%
\subsection{2-Variable Transient Growth with Dissipation present} \label{4.6}

\begin{figure}
\centering
\includegraphics[width=0.48\linewidth]{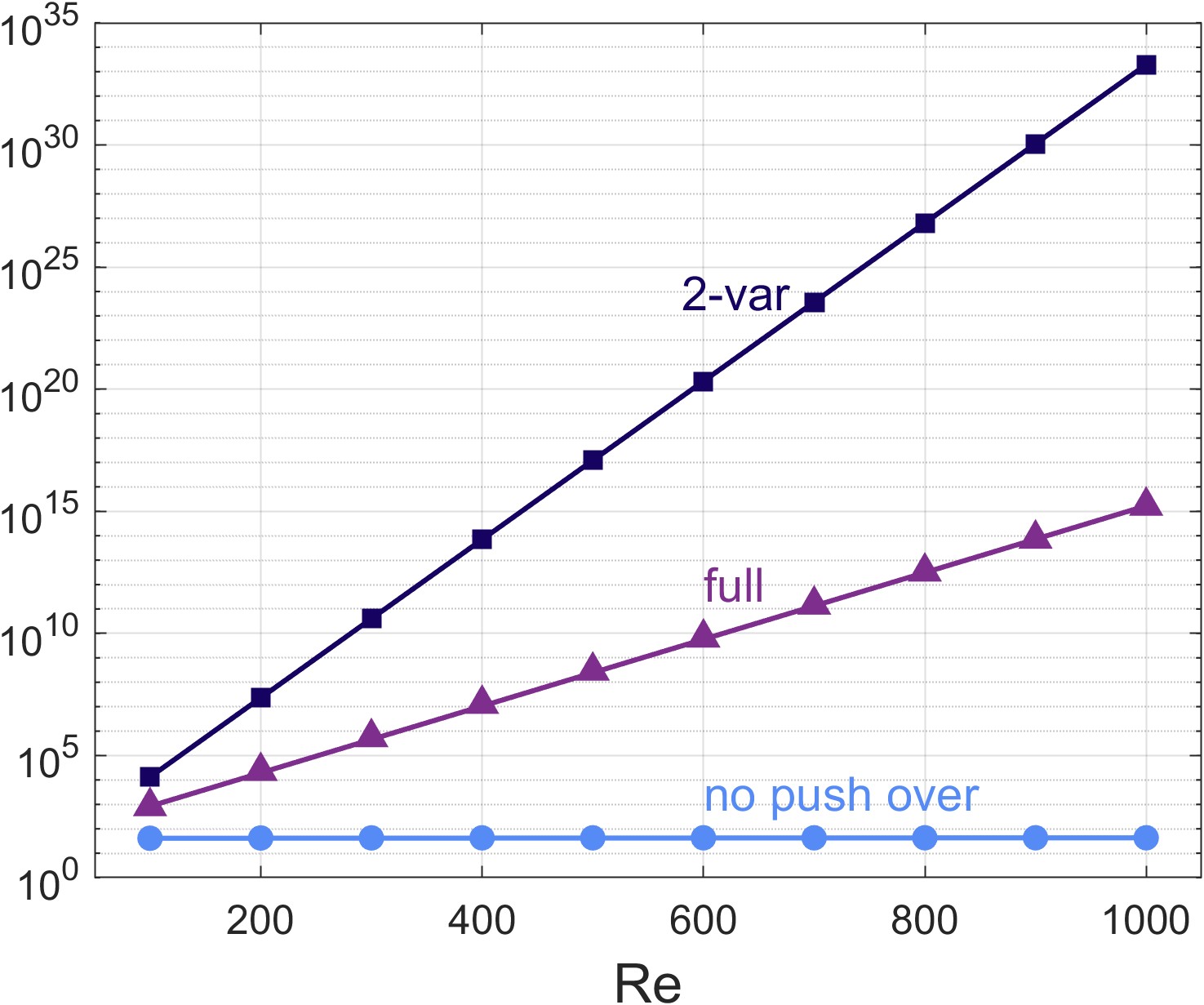}  
\includegraphics[width=0.47\linewidth]{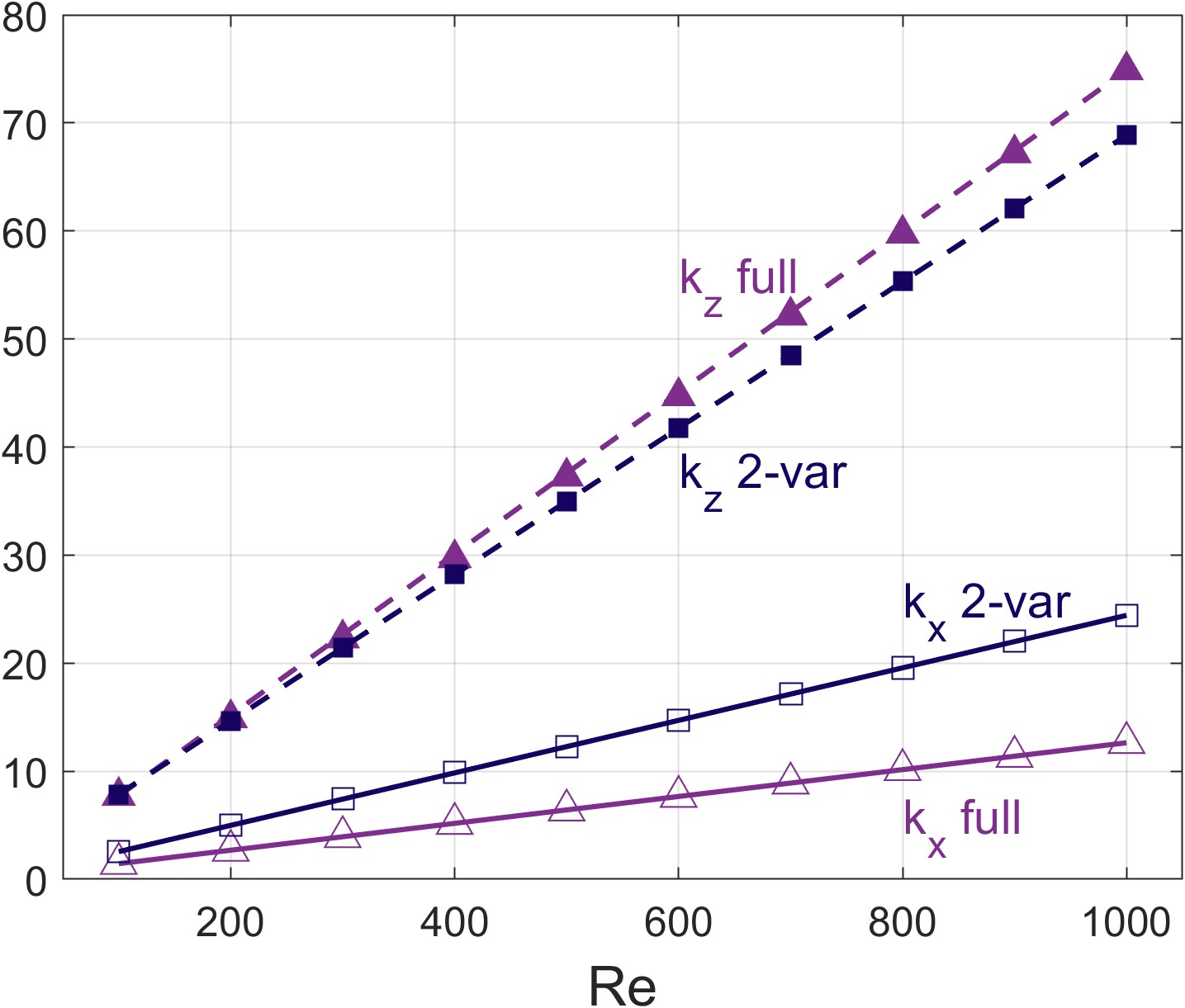}  
\caption{Left plot shows the global optimal gain $\goG$ as a function of $\Rey$ for $\beta=1$, $T=5$ using the full ($M=20$) system problem (purple triangles), the reduced 2-variable model (dark blue squares) and the full system with push-over removed (dull blue circles).
The lines drawn through the data are the straight lines between the extreme points of each data set indicating that the full system has the same $\goG \sim \mathrm{e}^{\alpha Re}$ behaviour as the 2-variable model albeit with a smaller $\alpha$. The push-over-less system does not have this exponential dependence. Right plot compares the optimal wavenumbers between the full system and the reduced 2-variable model (the symbols). Lines through the data (drawn as in the Left plot) indicate that all wavenumbers scale linearly with $\Rey$.}
\label{ReScale}
\end{figure}

Before studying the nonlinear consequences of the linear instability, we estimate the energy growth possible  in the 2-variable system with dissipation present. Because of the unlimited growth of the cross-stream wavenumber in the model, dissipation always eventually overpowers the instability to formally give only transient growth. For a general $k_x$, the time for the cessation of linear instability growth is given by the balance $\beta k_x \sim k_x^2T^2/\Rey$ so $T=O(\sqrt{\beta \Rey/k_x})$ and then a prediction for  the optimal growth for a {\em given} $k_x \lesssim O(\Rey)$  is
\beq
G \sim \mathrm{e}^{\sqrt{\alpha \beta^3 k_x Re} },
\label{largegrowth}
\eeq
where $\alpha=\tfrac{1}{2}(k_z^2-k_x^2)/(k_z^2+k_x^2)$ is $O(1)$. Formally, if $T\gtrsim O(1)$, this is maximized for $k_x =O(\Rey)$ beyond which diffusion dominates giving a massive gain of $O(\mathrm{e}^{\alpha_1 Re})$ ($\alpha_1$ being a $\Rey$-independent number). This prediction is confirmed in the 2-variable model {\em and} for the full system: see Figure \ref{ReScale}. This also shows good correspondence between the optimal wavenumbers in each system 
which all scale linearly with $\Rey$.
Figure \ref{ReScale} also confirms  that a push-over-less system has no such exponential growth.
 However, this overall optimal gain is not the one of practical interest as $k_z=O(\Rey)$ (figure \ref{ReScale} actually has $k_z \sim 0.075\Rey$ and $k_x \sim 0.012\Rey$) represents unrealistically large streak shear. If, instead $k_z=O(1)$ (and hence $k_x=O(1)$), the gain is reduced but still substantial at $O(\mathrm{e}^{\alpha_2 \sqrt{Re}})$ generated over $O(\sqrt{\Rey})$ `intermediate' times ($\alpha_2$ being another constant).
This exponential dependence of gain on $\sqrt{\Rey}$ is surprising given that the Orr and push-over mechanisms in isolation only give gains which scale like $\Rey^2$ and then only over `slow' times  of $O(\Rey)$ (e.g. see \S \ref{nostreak} and \S \ref{2DTG}).

%
%
%

%
%
\subsection{Nonlinear Feedback of the 2-Variable System} \label{4.7}

The linear instability in 2D Kolmogorov flow is known to be supercritical \citep{Sivashinsky85} and hence the first nonlinear feedback of the instability on the streak is to reduce its amplitude. As confirmation, the instability  here is
\beq
{\bf u}= \left[
\begin{array}{c}
\tfrac{2k_z}{h_1^2}\hat{\eta}_1(t) \sin k_z z \\
0\\
\tfrac{2ik_x}{h_1^2} \hat{\eta}_1(t) \cos k_z z+\tfrac{i }{k_x}\hat{\eta}_0(t) 
\end{array}
\right] 
e^{ i( k_x x+(1-k_x t)y) }+c.c.
\eeq
($c.c.$ indicating complex conjugate)  which, indeed, has the negative feedback  on the streak, 
\beq
\frac{k_x k_z}{2\pi^2}
\int^{2\pi/k_z}_0  \!\! \int^{2\pi/k_x}_0 (-{\bf u} \cdot \nabla u) \cos k_z z\, dx dz \, = \, -\frac{4 k_z^2}{k_x h_1^2}{\mathfrak Re} \! \left(\, i \hat{\eta}_0 \hat{\eta}_1^* \, \right) < 0
\eeq
as $\hat{\eta}_1 = i\hat{\eta}_0 \sqrt{k_z^4/k_x^4-1}/\sqrt{2}$
from (\ref{2vardiff_b}) ignoring diffusion ($^*$ indicates complex conjugate and ${\mathfrak Re}( \, )$ is the real part). So this 2-variable system does not re-energise the imposed streaks and moreover has no feedback on any streamwise rolls. To generate the latter, $\hat{v}_1$ needs to be reinstated - i.e. we need to examine the full  3-variable $M=1$ system - but at a cost of expecting less growth.

%
%
\begin{figure}
\centering
\includegraphics[width=0.3\textwidth, height=0.45\textwidth]{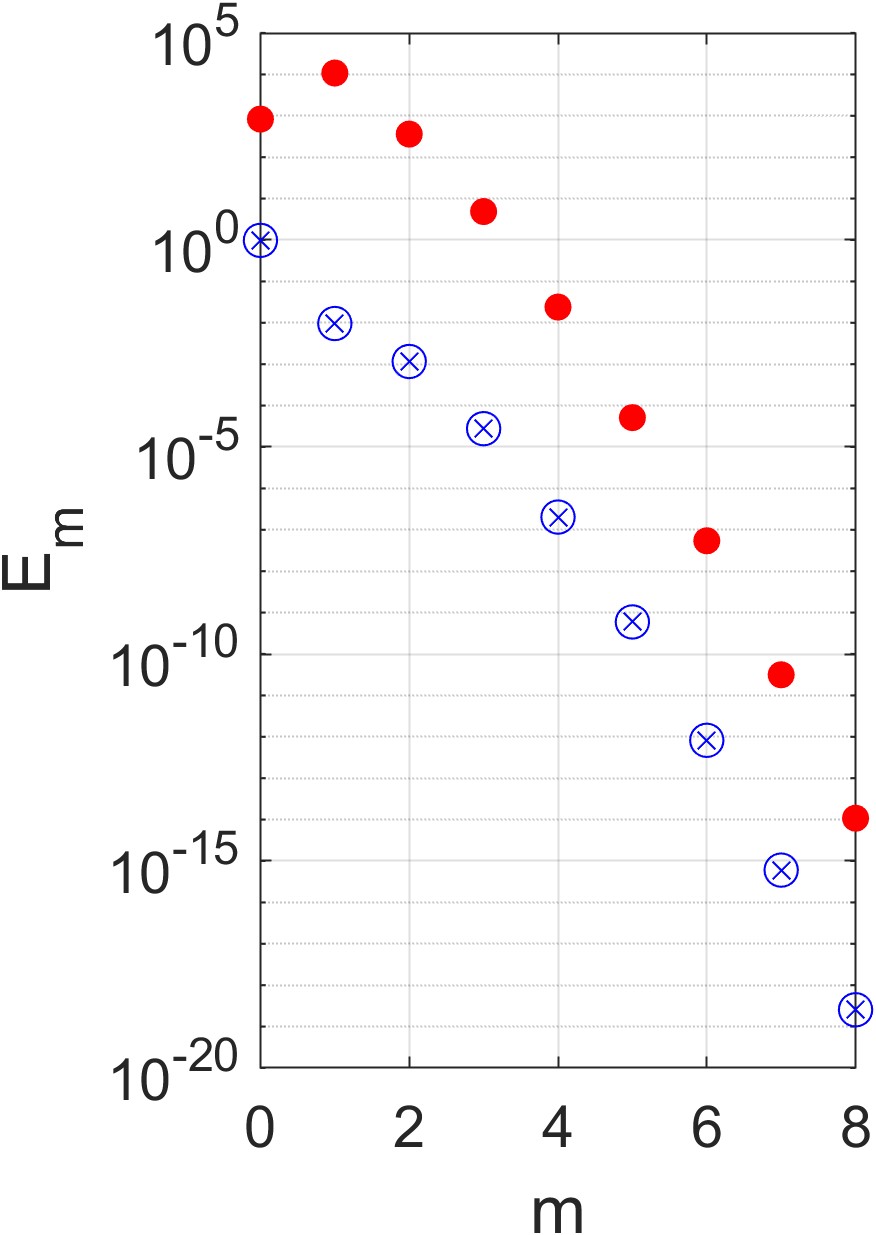}
\hspace{2mm}
\includegraphics[width=0.53\textwidth]{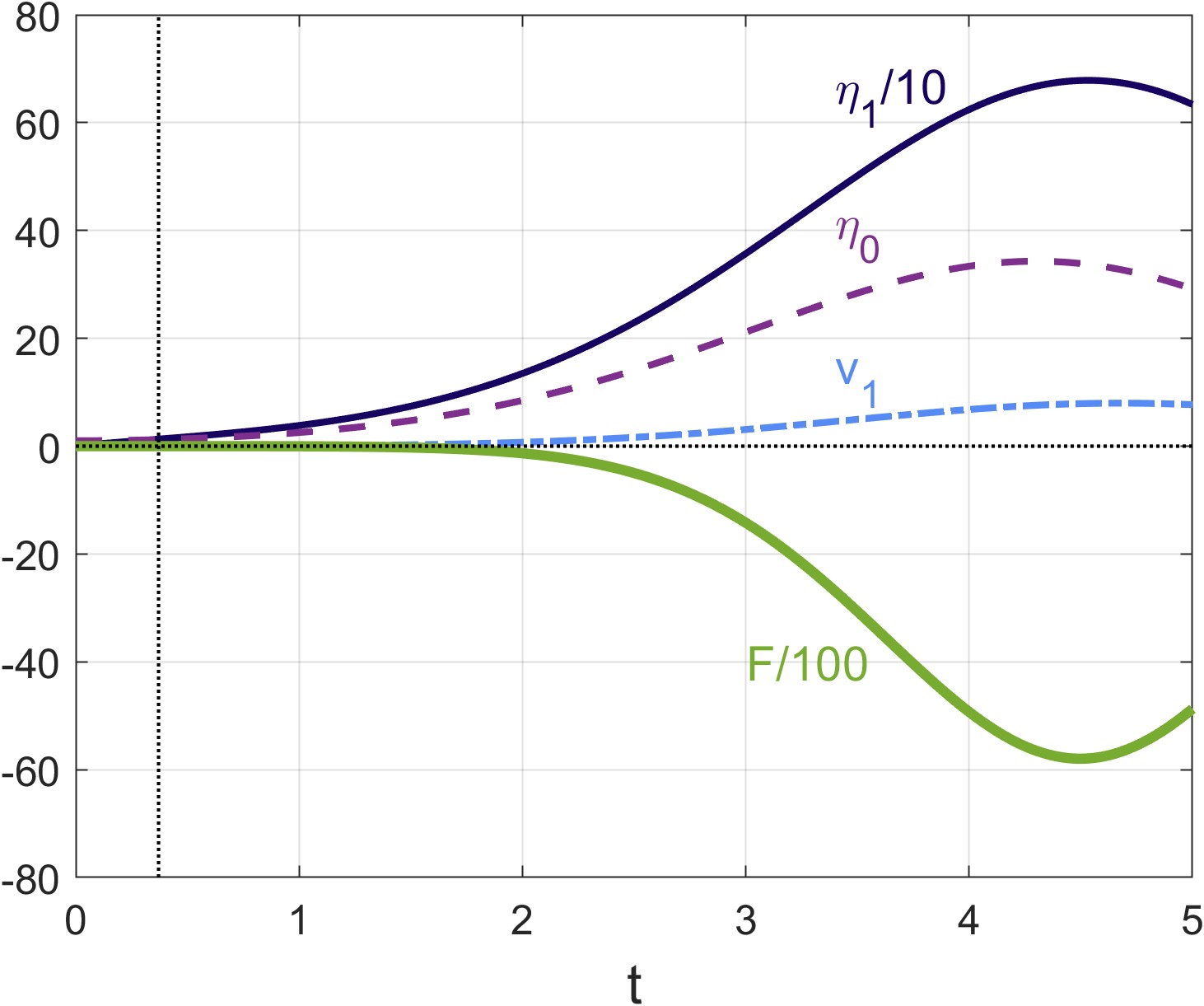}
\caption{Left: modal energy  $E_m:=\frac{1}{2h_m^2} (k_m^2 |\hat{v}_m|^2+|\hat{\eta}_m|^2)$ plotted over $m \geq 0$ ($E_{-m}=E_m$ due to symmetry) for the initial optimal (blue circled crosses) and final state (red filled circles) for $(Re,T,\beta)=(200,5,1)$ and $M=20$. This indicates why the $M=1$ truncation works so well: the initial optimal has its energy dominantly in the $m=0$ mode which shifts to $m=1$ by the final state. 
Right, the time evolution of the optimal for $M=1$ system and $(Re,T,\beta)=(200,5,1)$: solid dark blue uppermost line is $\eta_1/10$; dashed purple line is $\eta_0$ and the dash-dot pale blue line is $v_1$. The green solid line indicating negative values  is $F/100$ where $F$ is defined in (\ref{F}). The dotted vertical line at $t \approx 0.37$ indicates where $1-k_x t=0$, $k_x\approx 2.7$ and $k_z \approx 15$.
}
\label{evolution}
\end{figure}

%
%
\subsection{3-Variable Feedback onto Rolls} \label{4.8}

The $M=1$ linear system, (\ref{dv1})-(\ref{deta1}) has a symmetry that if $(\hat{\eta}_0, \hat{\eta}_1,\hat{v}_1)$ is a solution then so is $(\hat{\eta}_0^*, -\hat{\eta}_1^*,\hat{v}_1^*)$. This symmetry is adopted by optimal initial conditions so that the flow subsequently can then be assumed of the form $(\hat{\eta}_0, \hat{\eta}_1,\hat{v}_1)=(\eta_0(t),i\eta_1(t),v_1(t))$ where $\eta_0(t),\eta_1(t)$ and $v_1(t)$ are all real variables subsequently (this observation has already been used to produce equation (\ref{2varnodiff})\,). The evolution of these 3 optimal real variables is shown in Figure \ref{evolution} (right) for $(Re,T,\beta)=(200,5,1)$. The 3-variable flow field takes the form
\beq
{\bf u}= \left[
\begin{array}{c}
2i \left\{ -k_x(1-k_xt) v_1(t)+k_z \eta_1(t) \right\}\sin k_z z/h_1^2 \hspace{1.7cm}\\
\hspace{1.8cm}2 i v_1(t) \sin k_z z\\
2 \left\{ -k_z(1-k_xt)v_1(t)-k_x \eta_1(t) \right\}\cos k_z z/h_1^2+i\eta_0(t)/k_x 
\end{array}
\right] 
e^{ i( k_x x+(1-k_x t)y) }+c.c.
\label{M1flow}
\eeq
and the nonlinear driving of the cross-stream flow component $V(z,y){\bf \hat{y}}$ with the same spanwise structure relevant for lift up is
\beq
F:=\frac{k_x k_z}{2\pi^2}
\int^{2\pi/k_z}_0  \!\! \int^{2\pi/k_x}_0 (-{\bf u} \cdot \nabla v) \cos k_z z\, dx dz \, = \, 
-\frac{4 k_z}{k_x}\eta_0(t) v_1(t).
\label{F}
\eeq
For the representative parameters used in Figure \ref{evolution} (right), $\eta_0>0$ and $v_1>0$  and so the feedback $F$ is large and negative  (note that $F/100$ is actually plotted). This drives  secondary cross-stream velocities anti-correlated ($ \propto -\cos k_z z$) to the imposed streak field  which, through the lift up term $-U_y V \,{\bf \hat{x}}$ reinforces the existing streak field. Specifically, where the instability generates $V>0$, slower moving fluid from the basic shear is `lifted up'  to reinforce the slow streaks (where $\beta \cos k_z z <0$) and where $V<0$, corresponding faster-moving fluid is `pushed down' to reinforce the fast streaks (where $\beta \cos k_z z >0 $). Since the driven flow $V(y,z) {\bf \hat{y}}$ is synonymous with streamwise rolls, this demonstrates the potential for this streak instability to produce a sustaining cycle of rolls and streaks \citep{Hamilton95,  Waleffe97, JimenezPinelli99, FarrellIoannou12}.

\begin{figure}
\centering
\includegraphics[width=0.55\textwidth]{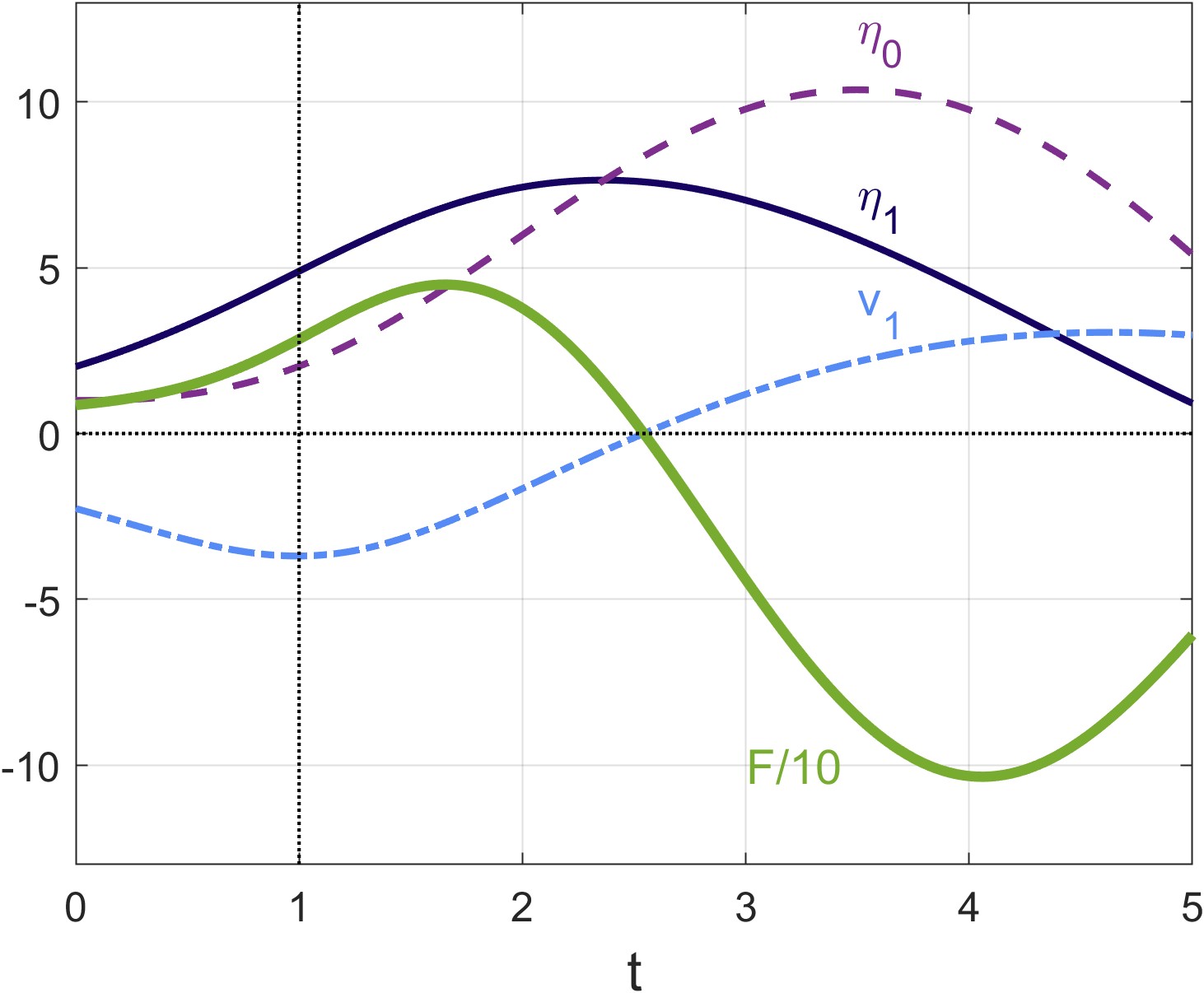}
\caption{The time evolution of the optimal for $M=1$ system and $(Re,T,\beta)=(200,5,1)$ with non-optimal $k_x=1$ and $k_z=0.95$ so a linear stable situation: solid dark blue line is $\eta_1$; dashed purple line is $\eta_0$ and the dash-dot pale blue line is $v_1$. The green solid line indicating largly negative values  is $F/10$ where $F$ is defined in (\ref{F}). The dotted vertical line indicates when $1-k_x t=0$.
}
\label{evolution1}
\end{figure}

Figure \ref{evolution1} confirms for non-optimal wavenumbers - $k_x =1$ and $k_z=0.95$ - with no instability present that even the transient growth acting alone provides good nonlinear feedback onto streamwise rolls.

%
%

%
%
\subsection{Why Lift Up Weakens Growth} \label{4.9}

Figure \ref{comp3} shows that removing $\hat{v}_1$ from the $M=1$ system considerably enhances the optimal growth. This is because both the $\hat{v}_1$-Orr term in equation (\ref{deta0}) and the lift-up term in (\ref{deta1}) weaken the instability mechanism which dominates the growth found here. Starting with Figure \ref{mechanism} (left) or equation (\ref{deta1}), the presence of the lift-up term works against the push over effect of $\hat{\eta}_0$. The $x$- and $z$-phases of $\hat{v}_1$ are the same  as that for $\hat{u}_1$ created by $\hat{\eta}_0$ (see (\ref{M1flow})\,) so $\hat{v}_1$ is either lifting up or pushing down the base shear flow to reduce the magnitude of $\hat{u}_1$. By this same reasoning, lift up can enhance the transient growth process if Orr dominates push over - i.e. there is no instability - but this scenario is never selected when maximising growth over wavenumber space.

In Figure \ref{mechanism} (right), the possibility of motion into or out of the $x$-$z$ plane of the streaks reduces the spanwise flow field $\hat{w}_1$ generated by the $\hat{u}_1$ field through continuity (the pressure field associated with the spatial gradients of $\hat{u}_1$ now can drive both $\hat{v}_1$ and $\hat{w}_1$ fields rather than just $\hat{w}_1$).  This reduced $\hat{w}_1$  then has weakened spatial gradients and so diminishes the streak Orr effect. In terms of equation (\ref{deta0}), this manifests itself as the $\hat{v}_1$-Orr term acting against the $\hat{\eta}_1$-Orr terms once $(1-k_xt)<0$ which happens early in the evolution (\,e.g. see Figure \ref{evolution} (right)\,).

%
%
\subsection{Previous work} \label{4.10}

The linear instability isolated here in equation (\ref{2varnodiff}) is inviscid in nature and, at least in terms of the streamwise  and cross-stream velocity components, is centred at the  spanwise inflexion points of the streaks where the streak shear is maximal. This suggests it is related to the linear instability originally observed by \cite{Swearingen87}. Against this, the spanwise flow components are maximal at the streak flow extrema, but there are no cross-stream inflexion points in this model and so no Kelvin-Helmholtz type instability  \citep[e.g. see figure 19b of ][]{Kline67}.

The model has revealed two transient growth mechanisms that can be in play for different wavenumber pairings. The one which is selected by optimising over the wavenumbers corresponds  to that found originally by  \cite{Sch-02} (e.g. their initial condition is very similar to the optimal conditions found here dominated by $\hat{\eta}_0$). There, the authors concentrated on the growth of the streamwise vorticity (e.g. see their figure 11 (right) and \S4.2) identifying what they called a `shearing' generation of vorticity due to the right hand side term in their equation (14a) and illustrated in their figure 16. This is a purely Orr term as there are no lift-up or push-over terms in the streamwise vorticity equation and  those authors attributed it predominantly to the base cross-stream shear (see just below (13) in \cite{Sch-02}). Here, we instead find that push over is the more important mechanism which has to dominate (streak) Orr to see this growth. When this happens, the 2-variable inviscid model is also linear unstable, that is, there are no wavenumber pairs for which this transient growth occurs with the flow stable (see Figure \ref{TG-LI}). Plausibly, introducing diffusion which stays bounded rather than growing steadily as here  would introduce a threshold for instability so recreating the possibility of a stable, transient growth scenario studied by \cite{Sch-02}. 

Finally, it is worth remarking that, for reasons of simplicity, we have identified the Orr mechanism only with advection terms  and these don't contribute to the perturbation energy equation. This means that there {\em must} be either lift up or push over operating  as well to actually get energy growth. The model studied here strongly suggests that push over is the key mechanism which is consistent with the findings of \cite{Hoepffner05}: their figure 6(a) shows that energy input from the streak field dominates that from the base shear.

\color{black}

%
%
\section{Discussion} \label{disc}

In this paper, motivated by recent numerical experiments on near-wall turbulence by \cite{Loz-21}, we have considered Kelvin's unbounded, constant-shear model augmented by a spatially-periodic spanwise shear to include streaks. This addition allows a little-studied transient growth mechanism - push-over - to be explored along with its interactions with the more familiar Orr and lift-up mechanisms. Consistent with the findings of \cite{Loz-21} and subsequent analysis by \cite{Mar-24}, this model clearly shows that the Orr and push-over mechanisms can combine to produce considerably enhanced transient growth for streaky base flows over that produced individually, and that lift-up is unimportant, actually tending to reduce peak growth (although, of course, lift-up is believed central for the roll-to-streak regenerative process).

The heart of this symbiotic interaction is laid bare in equation (\ref{2varnodiff}) which describes the evolution of a stripped down 2-variable version of the model.  This shows that for the optimal set of perturbation wavenumbers, there are actually two distinct mechanisms for growth:
1. a transient growth mechanism by which the spanwise-independent but streamwise-periodic cross-stream vorticity $\hat{\eta}_0$ generates spanwise-{\em dependent} streamwise-periodic cross-stream vorticity $\hat{\eta}_1$ involving push over and Orr processes, and 
2. a linear instability mechanism where $\hat{\eta}_1$ reciprocates by generating $\hat{\eta}_0$ via an Orr process in such a way as to give sustained asymptotic growth. 
Only when push over dominates Orr does the output of the growth - $\hat{\eta}_1$ - have the right spanwise phase to feedback on the input of the growth - $\hat{\eta}_0$ - to produce sustained asymptotic growth. In the model, wavenumbers invariably exist to achieve this and so very large growth occurs. 

Reinstating diffusion always eventually overpowers this growth due to the unlimited increase of the cross-shear wavenumber (an unfortunate feature of the model) but, in the meantime, huge growth, scaling like $\mathrm{e}^{\alpha_1 Re}$, can occur over `fast' $T=O(1)$ times. Even restricting the wavenumbers considered to be $O(1)$ appropriate for a turbulent boundary layer can produce growths scaling like $\mathrm{e}^{\alpha_2 \sqrt{Re}}$ over  `intermediate'  $T=O(\sqrt{Re})$ times (with $\alpha_1$ and $\alpha_2$ constants). This is  in marked contrast to the algebraic growth factors associated with Orr and lift-up, and now also shown here for push-over in isolation (see (\ref{2dPO3})), of gain $\sim \Rey^2$ over `slow' $T=O(\Rey)$ times. 

Reinstating the presence of a cross-shear velocity $\hat{v}_1$ is found to weaken the instability mechanism in two ways. Firstly, the reinstalled lift-up mechanism hampers the action of push-over in the transient growth mechanism. Secondly, there is an antagonistic $\hat{v}_1$-Orr term - third term on right hand side of (\ref{deta0}) - which hinders the generation of $\hat{\eta}_0$. This term grows with the cross-shear wavenumber and eventually curtails the instability completely (and, for the parameters studied, actually before diffusion). The presence of $\hat{v}_1$, however, is crucial in setting up the correct nonlinear feedback to generate the right type of streamwise rolls to re-energise the imposed streaks via lift up. Hence this model indicates the presence of the sustaining cycle in a simple shear flow.

With regards time scales, even just considering Kelvin's original model (no streaks) exposes the oversimplification of labelling mechanisms 
with one timescale based on what turns it off. Orr is considered `fast' and lift-up 'slow' yet both give the same
levels of growth at $T=O(1)$ - see (\ref{Orr_T_1}) and (\ref{Lift_up_T_1}) - and their  overall optimal gains are both achieved at $T=O(\Rey)$ and scale similarly with $\Rey^2$ - see (\ref{Orr_asym}) and (\ref{LiftUp_asym}). Just to complicate matters, the push-over mechanism behaves exactly equivalently - see (\ref{Push_T_1}) and (\ref{2dPO3}) - so there is no timescale puzzle: they all operate across inertial and viscous times. What perhaps is a surprise is that the  relevant Orr mechanism revealed here is based on streak advection  rather than the usual basic shear advection. 

Of course, the model treated here has its limitations. The streaks have no cross-shear structure (as there are no boundaries although \cite{Sch-02} comment that this is unimportant for the important sinuous modes - see their p67) so there are no cross-shear inflexion points. 
There is also the unlimited growth of the shearwise wavenumber so that diffusion eventually overpowers any apparent linear instability. 
Nevertheless, the model manages to clarify the transient growth mechanism originally found by \cite{Sch-02} (which is pushover dominated),  and also indicates how this transient growth  is related to the streak instability centred on the spanwise inflexion points observed by \cite{Swearingen87} (it forms one half of the loop). Using it, we have also been able to verify  that the correct nonlinear feedback occurs onto the right streamwise rolls to re-energise the assumed streaks. Finally, a connection has also been made to the well-known linear instability in 2D Kolmogorov flow.

\vspace{1cm}
\noindent
\backsection[Acknowledgements]{The authors are very grateful to two referees whose comments helped considerably improve the presentation of this work.}

\backsection[Funding]{W.O. acknowledges financial support from EPSRC in the form of a studentship.}

\backsection[Declaration of interests]{The authors report no conflict of interest.}

\backsection[Author ORCIDs]{

William Oxley, https://orcid.org/0000-0001-7434-0292;

Rich R. Kerswell, https://orcid.org/0000-0001-5460-5337.}

\appendix

\section*{Appendix: The Absence of Push-Over in the Evolution of $z$-independent Vorticity} 

This appendix explains the reason for the absence of any push-over contributions in the evolution equation (\ref{deta0}) for $\hat{\eta}_0$. This is not a trivial fact when using (\ref{Gveta2}) as a starting point as an Orr and a push-over term must be combined using continuity to arrive at the term involving $\partial v / \partial y$. The source of all the terms can be identified more clearly by starting with the original momentum equation given by (\ref{G1}) (where the terms associated with each physical mechanism are clear), and advancing towards equations (\ref{Gveta1}) and (\ref{Gveta2}) without making any simplifications or cancellations. The final two terms in (\ref{Gveta2}) are broken down as
\begin{equation}
-\overbrace{\beta k_z \sin{(k_zz)}\pard{u}{x}}^{\text{Orr}} -\overbrace{\beta k_z \sin{(k_zz)}\pard{w}{z}}^{\text{Push}} - \overbrace{\beta w k_z^2 \cos{(k_zz)}}^{\text{Push}},
    \label{AppA1}
\end{equation}
where the push-over contributions shown here are, in fact, the only push-over terms present in the evolution equation for $\eta$ (the advection term in (\ref{Gveta2}) is associated with Orr while the $\partial v/\partial z$ term is associated with lift-up).

To proceed towards equation (\ref{deta0}), the Kelvin modes in (\ref{KWs}) are used with $M=1$ along with the sinuous symmetry property discussed in section \ref{simpmodsec}. The variables can be written in a reduced manner, in particular $w=\left[ \hat{w}_0 + 2\cos{(k_zz})\hat{w}_1\right]\mathrm{e}^{\mathrm{i}\left[k_xx+(1-k_xt)y\right]}$. The terms labelled as push-over in (\ref{AppA1}) are then simplified to
\begin{equation}
2\beta k_z^2 \sin^2{(k_zz)}\hat{w}_1 - 2\beta k_z^2 \cos^2{(k_zz)}\hat{w}_1 - \beta k_z^2\cos{(k_zz)}\hat{w}_0,
    \label{AppA2}
\end{equation}
where the exponential multiplication factor is removed for clarity. The first two terms are merged to produce a quantity involving $\cos{(2k_zz)}$, which shows that there are no $z$-independent contributions found from the push-over mechanism; it will not contribute in the evolution equation for the $z$-independent component of vorticity, $\hat{\eta}_0$. This makes it clear that the $\hat{v}_1$ term in (\ref{deta0}) (that is subsequently ignored) is only of Orr origin, as is the $\hat{\eta}_1$ term that is retained.

The last term in (\ref{AppA2}) also produces the full push-over contribution to the $\hat{\eta}_1$ evolution equation given by (\ref{deta1}). The identification of Orr and lift-up is then made more simple as any remaining terms will be an Orr contribution if and only if they contain a factor of $\beta$.

\bibliography{resubmitted}
\bibliographystyle{jfm}

\end{document}